\documentclass[11pt,a4paper]{article}
\usepackage[a4paper,margin=28mm]{geometry}
\usepackage[T1]{fontenc}
\usepackage[utf8]{inputenc}
\usepackage[british]{babel}
\usepackage{amsmath,amssymb,amsthm}
\usepackage{booktabs}
\usepackage{longtable}
\usepackage{graphicx}
\usepackage{xcolor}
\usepackage{natbib}
\PassOptionsToPackage{hyphens}{url}
\usepackage[hidelinks]{hyperref}
\usepackage{xurl}

\graphicspath{{figures/}}
\theoremstyle{definition}
\newtheorem{definition}{Definition}[section]
\theoremstyle{plain}

\newcommand{\HFproviders}{36}
\newcommand{\HFoffers}{7249}
\newcommand{\HFmin}{1.69}
\newcommand{\HFmax}{10.00}
\newcommand{\HFmedian}{3.23}

\newcommand{\HFdays}{91}

\newcommand{\HFfirstdaylong}{5~July~2026}
\newcommand{\HFlastdaylong}{19~November~2026}
\newcommand{\GRPIlevel}{112.0}
\newcommand{\GRPIbase}{100.0}

\newcommand{\GRPIbasedatelong}{5~July~2026}

\newcommand{\GRPIlastdatelong}{19~November~2026}
\newcommand{\GRPIdays}{138}
\newcommand{\GRPIsigmaday}{0.426}
\newcommand{\GRPIsigmaTrading}{6.8}
\newcommand{\GRPIsigmaCalendar}{8.1}
\newcommand{\GRPImaxmove}{1.79}
\newcommand{\HorizonDays}{20}
\newcommand{\GRPIsigmaHorizon}{5}

\newcommand{\OCPIdays}{149}
\newcommand{\OCPIfirstdatelong}{23~June~2026}
\newcommand{\OCPIlastdatelong}{18~November~2026}
\newcommand{\OCPIsigmaday}{4.67}
\newcommand{\OCPIsigmaCalendar}{89}
\newcommand{\OCPIsigmaHorizon}{33}
\newcommand{\OCPIvrHorizon}{0.14}
\newcommand{\OCPIwindows}{7}
\newcommand{\Notecoupon}{19.86}
\newcommand{\Noteput}{15.78}

\newcommand{\NotecouponCalendar}{4.1}

\newcommand{\NoteputCalendar}{0.0084}
\newcommand{\NotecouponOCPIHorizon}{9.6}
\newcommand{\NoteputOCPIHorizon}{5.50}
\newcommand{\NotecouponOCPIDay}{31.4}
\newcommand{\NoteputOCPIDay}{27.36}
\newcommand{\DealsCompleted}{78.9}
\newcommand{\DealsCompletedN}{13}
\newcommand{\DealsTierA}{51.6}
\newcommand{\DealsTierAN}{12}

\newcommand{\BSstart}{23~August~2026}
\newcommand{\BSend}{18~November~2026}

\newcommand{\BSweeks}{12}
\newcommand{\BSheld}{70}
\newcommand{\BSgap}{18}

\newcommand{\BSdRHone}{+9}

\newcommand{\BSdPAnthropic}{$-$46}
\newcommand{\BSdQAnthropic}{+102}
\newcommand{\BSdQBAnthropic}{+185}

\newcommand{\BSdPOpenAI}{$-$11}

\newcommand{\BSdPGoogle}{+94}

\newcommand{\BSdPDeepSeek}{$-$62}

\newcommand{\CPvintage}{19~November~2026}
\newcommand{\CProws}{122}
\newcommand{\CPscored}{103}
\newcommand{\CPunscored}{10}
\newcommand{\CPuntil}{19~November~2026}

\newcommand{\CPfactorA}{1.0}

\newcommand{\CPfactorB}{1.6}

\newcommand{\CPfirstdateC}{7~August~2025}
\newcommand{\CPfirstpriceC}{3.44}
\newcommand{\CPfirstmodelC}{gpt-5}
\newcommand{\CPnowpriceC}{0.20}

\newcommand{\CPagopriceC}{3.44}
\newcommand{\CPfactorC}{17}
\newcommand{\CPcheaperdateC}{17~December~2025}
\newcommand{\CPcheaperpriceC}{1.13}
\newcommand{\CPcheaperfactorC}{5.6}

\newcommand{\CWdebtdate}{30~September~2026}
\newcommand{\CWdebttotal}{47.51}
\newcommand{\CWdebtdated}{27.86}
\newcommand{\CWdebtresidual}{19.64}
\newcommand{\CWdebtpct}{41.4}
\newcommand{\IRENdebtdate}{30~September~2026}
\newcommand{\IRENdebttotal}{10.04}

\newcommand{\IRENdebtpct}{64.0}
\newcommand{\StatusDate}{19~November 2026} 

\title{Markets for Compute and Inference:\\
       Structure, Financing, and Risk Transfer}
\author{Gabriel Borrageiro\\[4pt]
  {\normalsize Department of Computer Science, University College London}\\
  {\normalsize Gower Street, London WC1E 6BT, United Kingdom}\\[2pt]
  {\normalsize \href{mailto:g.borrageiro@ucl.ac.uk}{\texttt{g.borrageiro@ucl.ac.uk}}
   \quad ORCID \href{https://orcid.org/0000-0002-0063-7103}{0000-0002-0063-7103}}}
\date{19 November 2026}

\begin{document}
\maketitle

\begin{abstract}
Capacity to run large language models is rented by the accelerator-hour, and
their output is sold by the million tokens. Both goods now have published price
indices, an over-the-counter forward has traded on one, compute futures are
listed offshore, and CME and ICE have specified futures on rental indices that await the
United States regulator. We ask what in these two markets can be hedged,
by whom, and at what price.

Three findings stand out. First, the price that exchange contracts settle on
is not the price that drives operators' revenue. The contracts on CME and ICE reference
published short-dated rental rates, while most revenue, and with it credit
quality, comes from private multi-year contracts whose prices are largely
undisclosed. Second, the token price and the rental rate are linked by an
identity through throughput and margin, but they are different exposures, so a
hedge on the rental rate does not hedge the cost of inference. Third, the demand for hedging already sits in credit
agreements, as the years by which operators' debt outlives their contracted
revenue. Filings cannot measure that gap, because debt is scheduled year by year while
contracted revenue rarely is, and the one
loan that prices it settles the risk by cash prepayment against a model only
the parties see.

We design a reverse convertible note through which leveraged operators could
buy protection against falling rental rates. Its coupon depends steeply on the
annualised volatility of the rental rate, which the two public indices put
between about \GRPIsigmaHorizon\% and \OCPIsigmaCalendar\% according to how each
is built and the horizon measured; across that range the fair one-year coupon
runs from \NotecouponCalendar\% to \NotecouponOCPIDay\%.
\end{abstract}

\tableofcontents
\clearpage

\section{Introduction}
\label{sec:introduction}

Capacity to run large language models is rented by the accelerator-hour, and
the output of those models is sold by the million tokens. Both have become
priced goods with published indices. An over-the-counter trade has been
executed against one of those indices, and compute futures are listed
offshore. CME and ICE have
specified monthly futures on rental indices, Nodal Exchange and Architect have
announced their own, and the Commodity Futures Trading Commission has asked
publicly how such contracts should be listed. We ask what in these two markets
can be hedged, by whom and at what price, and what can at present only be
assumed.

The question is harder than its counterpart in an established commodity, for
two reasons. The first is that capacity cannot be held. In a market for a
storable commodity, or for currency or a bond, a forward price is tied to spot
by a buy-and-hold trade: buy now, finance the position and deliver later. An
hour of accelerator time used today cannot be delivered next month, so no such
trade exists, and the forward curve is set by expectations and risk premia
rather than by spot plus the cost of carry. The second reason is that the visible price is not the one that matters.
In most commodity markets the liquid, visible, short-dated price is the
benchmark off which long-dated bilateral deals are priced. Here the multi-year
committed contracts are the economics, because they are what lenders
underwrite, yet they are bilateral and largely undisclosed, while the visible
hourly rate cards set a small share of operators' revenue. The contracts CME and ICE have
specified settle on that visible rental rate. Tokens
can be bought forward only physically, and no option on either good is listed.

We make five contributions. First, we describe the physical market, the venues,
the reference indices and the exchange contracts, and record the status of each
contract as at \StatusDate{}. Second, we state the identity linking
the token price to the rental rate through throughput and margin, show that a
hedge on one is not a hedge on the other, and read the identity on Ornn's
indices of both prices, which show the two moving apart
(Figure~\ref{fig:bridge-series}). Third, from filings we document the debt
raised against capacity revenue and show that hedging demand is the gap between
when operators' debt falls due and when their contracted revenue runs out. That
gap cannot be measured from published information, because operators schedule debt year by year and, with few exceptions, report
contracted revenue in coarse bands or as a single total; one credit
agreement settles it by cash prepayment against a model that is not public.
Fourth, we specify a process for the rental rate and show why it cannot yet be
estimated: the two rental indices freely available to us, one built from rate
cards and one from transactions, disagree about its volatility by a factor of about seven to
eleven, according to the horizon measured. Fifth, we design a reverse convertible note through which leveraged
operators could buy protection against a fall in rental rates, and show that
its coupon depends steeply on that unmeasured volatility.

Prices, index levels, filings and model scores are collected from public
sources, every figure and table is generated from those records, and the data
end on \HFlastdaylong{}.

Section~\ref{sec:notation} fixes notation. Section~\ref{sec:landscape}
describes supply and demand, Section~\ref{sec:instruments} the instruments,
and Section~\ref{sec:compute-and-inference} the relationship between the two
goods. Section~\ref{sec:price-formation} turns to price formation,
Section~\ref{sec:deals} to financing, and Section~\ref{sec:risk-transfer} to
the demand for risk transfer. Section~\ref{sec:products} sets out the note and
Section~\ref{sec:conclusion} concludes.

\section{Notation and definitions}
\label{sec:notation}

This section collects the notation used across the whole document. Sections
\ref{sec:price-formation} and \ref{sec:products} introduce further quantities
local to their own arguments.

\subsection{The two goods}

An \emph{accelerator} is a processor built for the dense parallel arithmetic
that neural networks require: a graphics processing unit or a purpose-built
equivalent. The NVIDIA H100 and B200 are the models on which CME and ICE have
written their compute futures (Section~\ref{sec:listed}). The accelerator is
the unit of capacity the industry buys, installs, finances and rents, and we
prefer the word to ``GPU'' because several of the devices in question are not
graphics processors.

A \emph{frontier laboratory} is a firm that trains and sells access to
general-purpose models at the leading edge of capability. Frontier
laboratories are the largest single buyers of compute and the sellers of most
inference; they appear throughout as buyers rather than sellers of capacity.

\begin{definition}[Compute]
\emph{Compute} is the rental of accelerated hardware for a period of time. The
buyer obtains the exclusive use of a stated number of accelerators of a stated
model, in a stated location, for a stated interval. Compute is an input good:
it is consumed to produce something else.
\end{definition}

\begin{definition}[Inference]
\emph{Inference} is the sale of model output. The buyer submits text or other
data to a trained model and pays for the output returned, metered by token
count. Inference is a finished good: it is what compute is consumed to
produce.
\end{definition}

Neither good can be bought today and held for later. An accelerator-hour that
passes unused is gone, and so is an idle inference endpoint. For a storable
commodity, a currency or a bond, a forward price is tied to the spot price by
exactly that trade: buy the good now and carry it to delivery, paying the cost
of financing and storage and earning any yield it provides. Without it, the
forward curve still has a shape, set by expected supply and demand, by a risk
premium and by speculative positioning, but nothing pins it to today's price.
Section~\ref{sec:carry} draws the consequences.

\subsection{Quantities}

We index calendar time by $t$ and write $T$ for the settlement or maturity date
of a contract. Two clocks are in use and we state which applies throughout: the
price dynamics of Section~\ref{sec:price-formation} are written in days,
because that is the sampling frequency of the underlying indices, while the
contract and discounting arithmetic of Section~\ref{sec:products} is written in
years, because that is how rates and coupons are quoted. Where both appear, 365
days is one year.

An \emph{accelerator model} is a specific piece of hardware, such as the
NVIDIA H100 or B200. We index accelerator models by $g$.

The \emph{rental rate} is the price paid to rent one accelerator for one hour.
It is the observable price of compute, and it is what capacity indices and
listed futures settle against. We write it $R^{(g)}_t$, in US dollars per
accelerator-hour.

The \emph{futures price} is the price agreed today for rental delivered in a
future settlement month. It is the quantity a hedger transacts in. We write it
$F^{(g)}_t(T)$, in dollars per accelerator-hour.

A \emph{language model} is a specific trained model sold through an inference
endpoint. We index language models by $m$.

The \emph{token price} is the price paid for one million tokens of model
output. It is the observable price of inference. We write it $P^{(m)}_t$, in
dollars per million tokens.

\emph{Throughput efficiency} is the quantity of output a model produces from
one accelerator-hour. It is the conversion factor between the two goods, and
improvements in it are one of the two mechanisms by which inference can become
cheaper while compute does not, the other being a lower margin. We write it $\varepsilon^{(m)}_t$, in millions of
tokens per accelerator-hour.

The \emph{margin} is the fraction by which a provider's token price exceeds
the rental cost of the accelerator-hours that produce the tokens. It is a
markup on that cost, not a share of revenue, and it absorbs every other cost of
serving a model as well as profit. It is dimensionless and we write it
$\mu^{(m)}_t$.

\emph{Fleet utilisation} is the fraction of installed accelerators earning
revenue at a given moment. It is dimensionless, lies in $[0,1]$, and we write
it $u^{(g)}_t$.

\subsection{The bridge between the two prices}

With these terms defined, the relationship between the price of inference
and the price of compute is an accounting identity rather than a model, since
it defines the margin:
\begin{equation}
\label{eq:bridge}
P^{(m)}_t \;=\; \frac{R^{(g)}_t}{\varepsilon^{(m)}_t}\,\bigl(1 + \mu^{(m)}_t\bigr).
\end{equation}
The units reconcile: dollars per accelerator-hour divided by millions of
tokens per accelerator-hour gives dollars per million tokens, and the margin
term is dimensionless.

Equation~\eqref{eq:bridge} is the reason the two markets require separate
instruments. A rise in $\varepsilon^{(m)}_t$, an improvement in throughput,
reduces $P^{(m)}_t$ with $R^{(g)}_t$ unchanged.
A capacity glut reduces $R^{(g)}_t$ while providers hold $P^{(m)}_t$ and take
the difference as margin. Hedging one exposure does not hedge the other.

\subsection{Note and hedge quantities}

The remaining symbols describe one specific instrument and its hedge. The instrument is a \emph{reverse convertible note}: a
bond paying an enhanced coupon, whose holder is repaid in full unless a
reference price finishes below an agreed level, the strike, in which case they
absorb the shortfall. Section~\ref{sec:products} sets out why that structure fits this
market and who would take each side; here we fix only the vocabulary.

The \emph{strike} is the rental level below which the holder ceases to receive
principal in full. We write it $K$, in dollars per accelerator-hour.

The \emph{coupon rate} is the rate paid to the note holder over the term of the
note, written $c$. For a one-year note it coincides with the annualised rate;
for any other tenor the two differ and we annualise explicitly. The
\emph{principal} is the amount subscribed, written $N$, in dollars.

The \emph{hedge ratio} is the number of units of a hedging instrument held per
unit of exposure. We write it $h_t$, dropping the
subscript to $h^{*}$ where we mean specifically the variance-minimising ratio.

\emph{Residual variance} is the variance of the hedged position that survives
the hedge. It measures the risk a hedger still carries, and we write it $\Sigma_t$, dropping the subscript where
the statement does not depend on time.

\subsection{Process parameters}

Where we later model $R^{(g)}_t$ as a mean-reverting process with jumps, the
parameters carry their conventional meanings: $\kappa$ is the rate of mean
reversion, per day; $\theta$ is the long-run mean level of the \emph{logarithm}
of the rental rate, so that $e^{\theta}$ carries the units of dollars per
accelerator-hour; $\sigma$ is the diffusive volatility, per square root of a
day; $\lambda$ is the jump arrival intensity, per day; and $J$ denotes the jump
size, measured in log points. Where a declining trend is also present, as it is
in Section~\ref{sec:price-formation}, $\theta$ is the reversion target rather
than the asymptotic level: the process lags a falling target and settles a
fixed distance above it. We defer the
process statement itself to Section~\ref{sec:price-formation}, where it is
first used.

\section{The landscape: supply and demand}
\label{sec:landscape}

Two things must be true before a commodity market forms. The good must be
produced by parties whose revenue falls when its price falls, and consumed by
parties whose costs rise when its price rises. Compute satisfies both, and
increasingly so, because the two sides are committed on opposite terms: supply
is built for decades and financed with fixed obligations, while the demand
that ultimately pays for it, inference bought by the token, is uncommitted and
reprices in weeks.

\subsection{Supply}

Installed capacity sits in three tiers that behave differently under price
stress.

A \emph{hyperscaler} is a firm operating data centres at a scale set by its
own product demand rather than by what it can let to others; Alphabet, Amazon,
Meta, Microsoft and Oracle are the conventional list. They build primarily for
their own workloads and rent the surplus. Oracle apart, which borrows heavily
to build (Section~\ref{sec:deals}), they fund their accelerator fleets from
operating cash flow, so a fall in $R^{(g)}_t$ compresses a hyperscaler's margin
without threatening the firm. An operator whose only business is letting
capacity has no such latitude, because its rate must cover debt service.

The independent operators, conventionally called neoclouds, rent capacity as
their only business. Synergy Research Group put the segment above
\$25\,billion in 2025, growing at 223\% year on year in the fourth quarter,
and forecast it approaching \$400\,billion by 2031 at a compound rate near
58\%~\citep{synergy2026neocloud}. These operators matter to us out of proportion
to their size, because they are the participants whose debt service is fixed
while their revenue floats. They are the natural buyers of downside
protection. Section~\ref{sec:deals} sets out the financing raised across the
sector.

Beneath both sits wholesale colocation: shells, power and cooling leased to
whoever installs the accelerators. This tier is financed like real estate and
transmits price stress on a slower cycle.

Which listed firms depend most on this ecosystem has been mapped from supplier
data. Sona Asset Management put the share of revenue dependent on artificial
intelligence at approximately 87\% for Applied Digital, 80\% for CoreWeave, 58\%
for SK Hynix, 52\% for NVIDIA and 46\% for TSMC~\citep{ftav2026sona}, read from
a chart published without a data table and so approximate to about a
percentage point. Two cautions attach. The measure is dependence on the
ecosystem as a whole, not exposure to rental rates: SK Hynix's runs through
memory sales and TSMC's through the volume of chips ordered. And a revenue
share is not a price sensitivity: fixed-price contracts, prices that already
discount a downturn and leverage can all make two firms with the same
dependence respond very differently to a fall in the rental rate. The ordering
is still a better screen for hedge candidates than intuition, which favours
famous names and misses firms such as Applied Digital, and
Section~\ref{sec:products} uses it only as a screen.

\subsection{The binding constraint is power}

Accelerators can be manufactured faster than they can be energised. McKinsey
projects total data-centre capacity demand rising from 82.3\,GW in 2025 to
219.0\,GW in 2030, a compound rate near 22\%~\citep{mckinsey2026workloads}. Delivering that requires grid
interconnection, transmission and generation on timescales set by utilities
and regulators rather than by semiconductor lead times.

The consequence for price formation is specific. When capacity is bounded by
energisation rather than by silicon, supply cannot respond to a price spike
within the horizon over which the spike occurs. That is the condition under
which spot prices become sharply and asymmetrically volatile, and it is the
reason Section~\ref{sec:price-formation} reaches for a jump process rather
than a diffusion.

\subsection{Demand, and where it is moving}

Demand divides into training, which consumes capacity in large scheduled
blocks, and inference, which consumes it continuously in proportion to end-user
activity. The mix is shifting toward the latter. On McKinsey's
projection, inference reaches 93.3\,GW by 2030 against 62.2\,GW for training
(more than 40\% of all data-centre demand against roughly 30\%), growing
at 35\% a year while training grows at 22\%. The remaining 63.5\,GW is
conventional computing that does not serve artificial intelligence, growing at
11\% a year.

The leg of demand that is growing fastest is thus the leg with no listed
instrument. Every
exchange contract on compute described in Section~\ref{sec:instruments} is written on
accelerator capacity, and none on the token price $P^{(m)}_t$, for reasons
Section~\ref{sec:token-leg} sets out.

\subsection{Why the two sides do not net out}

Supply is built for decades and financed with fixed obligations. The customer
contracts that support it run for a few years, and the final demand behind
those contracts, inference bought by the token, is not committed at all.
Between them sit the participants holding the mismatch: operators who have
signed for power and hardware against revenue they have not yet contracted. That mismatch is the
exposure a market in risk transfer would exist to carry.

\section{What is traded, and what each contract means}
\label{sec:instruments}

We describe the instruments in order of how directly they touch the two
goods: the physical rental market, the venues and indices built on it, and the
exchange contracts that settle against those indices.

\subsection{The physical market}
\label{sec:physical}

Capacity is sold in three shapes. \emph{On-demand} rental is billed by the
hour with no commitment, and it is the shape that published rate cards quote.
\emph{Committed} or reserved capacity is contracted for a term, from months to
several years, at a discount to on-demand; the large agreements with frontier
laboratories sit here, and it is what lenders underwrite. \emph{Interruptible}
capacity is sold at a deep discount and reclaimed when a higher-paying workload
arrives.

Only the first is transparently priced, and this is the central data problem
of the market. Committed rates are bilateral and are disclosed, if at all,
through the filings of whichever party must disclose them. In most commodity
markets the benchmark is the liquid, visible, short-dated price, and long-dated
bilateral deals are priced off it. Here the relation runs the other way: the
long-dated bilateral deals are the economics, because they are what the lender
underwrote, and the visible price is a thin residual. The prices that determine
credit quality are the least visible, and the prices that are visible
determine the smallest share of revenue. The indices behind the CME and ICE
contracts are built on the visible part, and so measure the part of the market
that matters least to solvency. Compute Desk's indices, on which Nodal's
announced contract and Architect's pending ones would settle, are a partial exception: they quote one-year
reserved rates for US neocloud capacity, blending completed transactions with
published list prices, both adjusted to that term and to a reference
accelerator in ways the methodology does not specify~\citep{computedesk2026method}.

\subsection{Venues away from an exchange}

One private venue and one dealer already trade compute. Compute Exchange
describes itself as an institutional marketplace in which identity-verified
counterparties trade physical accelerators, reserved cloud capacity and forward
contracts for future-dated delivery at a locked price, with quotes solicited by
auction across providers~\citep{computeexchange2026}. It also offers token
forwards: a buyer commits to a volume of inference tokens on an open-weight
model at a per-token rate locked for up to six months, and draws against it
over the term. Volume is measured in standardised units that weight an
uncached input token at 1.00, a cached input token at 0.20 and an output token
at 4.21~\citep{computeexchange2026tokens}. These are physically delivered
forward markets, one in compute and one in tokens, and they are where a
physical hedge would be executed. They are also young and privately run: the
only volumes reported are the operator's own figures for requests, not trades,
we have not transacted on the venue, and nothing about its scale or liquidity
can be checked from outside.

A derivative on compute has also traded over the counter. In May 2026 FalconX
executed what it describes as the first over-the-counter compute forward
trade, referencing Ornn's H100 index and settled against it, with one named
counterparty; size, tenor and terms were not disclosed~\citep{falconx2026otc}.
One trade is not a market, but it shows that the documentation, fallback and
valuation questions a first trade must answer have been answered once.

\subsection{The reference indices}

Derivatives settle against published indices rather than against any one
provider's rate, and the three indices that matter here are published by small
firms whose view of the market is partial. Two of them nonetheless matter for
the exchange contracts of Section~\ref{sec:listed}: CME's contracts settle on
Silicon Data's index and ICE's on Ornn's.

Silicon Data publishes daily rental indices for individual accelerator models,
normalised for specification, rental term and geography across on-demand,
interruptible and reserved offers; the normalisation is described only in
words, and the data are sold by subscription~\citep{silicondata2026index}. Ornn
builds its index only from prices actually paid under running rentals, as a
volume-weighted, trimmed mean over rolling one-hour windows, and leaves out
quotes and reserved contracts~\citep{ornn2026ocpi}. It also publishes a token
price index for each of four laboratories, weighted by transacted volume, for
which it publishes no methodology~\citep{techflow2026otpi}. A third index, from
GPU Rental Prices, is free, settles nothing and covers only posted rate cards.
It is chain-linked: each day's change compares each listed configuration with
its own recent price, so a provider entering or leaving cannot move it, and it
stood at \GRPIlevel{} on \GRPIlastdatelong{} against a base of \GRPIbase{} on
\GRPIbasedatelong{}~\citep{gpurentalprices2026}. This index, Ornn's public
window and our own price collection (Section~\ref{sec:compute-and-inference})
are the only rental series available to us without a subscription.

None of these indices is final once published. On 26~October 2026 GPU Rental
Prices restated every level since its base, replacing an arithmetic mean of
family medians, which drifted upward whenever a price moved up and back. On
that day the earlier method read 122.96 and the restated index
109.72~\citep{gpurentalprices2026}. Ornn's index revises too: between our
captures of 23~September and 10~November 2026, 143 of the 172 daily values the
two captures share changed, by up to 5.6\%. A contract settling on a published
index therefore needs a stated rule for revisions and for changes of formula.

The two constructions are open to opposite manipulations. A posted-price index
can be moved without trading, by revising a rate card on capacity the provider
barely sells, at almost no cost. A transaction index can be moved only by
trading, but in a thin market a loss-making trade large enough to move it may
be small beside a derivative position settled on it. The first needs volume
weighting and staleness filters; the second needs volume thresholds, caps on
any participant's share and a settlement window long enough that moving it is
expensive.

\subsection{The exchange contracts}
\label{sec:listed}

Table~\ref{tab:contracts} sets out what exchanges have filed, announced or
listed. The CME and ICE columns are taken from each exchange's own contract
specification; the Nodal and AX columns rest on each venue's description of its
own product.

\begin{table}[htbp]
\centering
\caption{Exchange contracts on compute, as at \StatusDate{}. None settles on
a token price, and none is an option.}
\label{tab:contracts}
\small
\begin{tabular}{@{}lllll@{}}
\toprule
 & \textbf{CME} & \textbf{ICE} & \textbf{Nodal} & \textbf{Architect AX} \\
\midrule
Product & GPU1, GPU2 & HPR, BKL & Compute futures & Compute futures \\
Reference & Silicon Data & Ornn & Compute Desk & not disclosed \\
          & daily indices & index & daily indices &  \\
Accelerators & H100, B200 & H100, B200 & Hopper, & H100, H200, \\
             &            &            & Blackwell & B200, B300 \\
Contract unit & 730 acc.-hours & Hours in month & not disclosed & not disclosed \\
Settlement & Monthly average & Monthly average & Financial & not disclosed \\
           & of daily index & of daily average &  &  \\
Listings & 36 monthly & Up to 24 monthly & not disclosed & Quarterly to \\
         &            &                  &               & Mar 2027 \\
Status & Under review & No listing date & Announced & Listed \\
       & by the CFTC & set & 3 Sep 2026 & (Bermuda) \\
\bottomrule
\end{tabular}
\end{table}

CME's contracts are the most fully specified. Each covers 730
accelerator-hours, one month of a single accelerator, and settles in cash
against Silicon Data's daily index for non-hyperscaler on-demand capacity, with
monthly listings out to 36 months and a tick value of
\$7.30~\citep{cme2026compute}. They were filed for the regulator's
\emph{approval}, rather than self-certified, so they cannot take effect by
default~\citep{nymex2026submission}. The Commodity Futures Trading Commission
(CFTC) extended its review, citing novel or complex issues and its request for
comment on compute derivatives, which closed on 20~October
2026~\citep{cftc2026extension,cftc2026rfc}; NYMEX has since consented to extend
the review of both contracts to 8~January 2027~\citep{cftc2026records}. ICE's
contracts, on Ornn's index, have no announced listing date, and ICE publishes
hypothetical daily settlement prices~\citep{ice2026notice}. Nodal Exchange has
announced futures on Compute Desk's indices for 2026, subject to regulatory
compliance~\citep{nodal2026compute}. Architect, which acquired a United States
exchange licence, lists metals and currency futures there and has announced
futures and options on compute with an exchange-for-physical link, through
which a futures position can be swapped for the underlying capacity; its
compute futures, cash-settled on Compute Desk's indices, await
approval~\citep{architect2026dcm,architect2026exchange}. The compute futures
listed on an exchange are offshore: Architect's Bermuda venue, AX, lists
quarterly futures on H100, H200, B200 and B300 capacity to March 2027, names no
reference index for them and publishes no volume~\citep{architect2026ax}.

Both exchange specifications settle on a monthly average rather than a single
fixing, which makes these average-price futures, the futures counterpart of an
Asian option. Averaging has two consequences. It raises the cost of
manipulating the settlement roughly in proportion to the number of days
averaged. And it changes the hedge: once $k$ of the $n$ days in the month have
fixed, only $n-k$ remain uncertain, so a position held into the delivery month
loses its sensitivity to the rental rate linearly as the average accumulates.

If they list, the CME and ICE contracts could draw liquidity that none of the
venues above can, because the banks, funds and commodity firms that would trade
them already clear there. Until then, statements in this document about what
the listed market offers apply from the first day a contract trades.

\subsection{The inference leg}
\label{sec:token-leg}

No cash-settled contract on a token price is listed or traded, and the reason
is the good itself. A contract on a named model loses its hedgers as the model
is superseded, which now happens within months, and its successor may be
cheaper or dearer per token. A contract on a capability threshold avoids that,
but needs a benchmark whose scale is itself revised as models improve
(Section~\ref{sec:fixed-capability}). Prices are also set by a few
laboratories, and the one published token index has no methodology. A design
in the literature takes the threshold route: it defines a standard token as
output from a model meeting stated benchmark thresholds, sizes a lot at one
million such tokens, and settles in cash against a volume-weighted token price
index in which no provider's weight exceeds 30\%~\citep{tokenfutures2026}. The
one traded alternative is the physically delivered token forward described
above.

\subsection{What remains uncovered}

Four exposures would have no instrument on CME or ICE even if their contracts
traded. Accelerator models other than H100 and B200 are hedged only by proxy.
Committed, interruptible and regional capacity carry \emph{basis}, the gap
between the price of the exposure a hedger holds and the price the contract
settles on, which Section~\ref{sec:compute-and-inference} takes up. The token
leg has only a physical forward on open-weight models. And with no option
listed anywhere, any position with a convex payoff must be warehoused or hedged
through instruments on other underlyings, which is where
Section~\ref{sec:products} begins.

\section{Compute and inference}
\label{sec:compute-and-inference}

Section~\ref{sec:notation} defined the two goods. Here we set out the
properties that determine how they can be priced and hedged, and why an
instrument written on one does not cover exposure to the other.

\subsection{Without a carry trade, the forward curve is not pinned to spot}
\label{sec:carry}

For a storable commodity, a currency or a bond, the forward price is held near
the spot price by a carry trade: buy the good today, finance and hold it, and
deliver it later. The forward therefore equals spot grown at the cost of
financing and storage, less any yield or convenience the holder
earns~\citep[ch.~5]{hull2015}. Neither compute nor inference admits the trade,
because an accelerator-hour cannot be bought today and held until next month.

The forward curve still has a shape. It reflects the expected future rental
rate, a risk premium that clears the imbalance between hedgers and
speculators at each maturity, and speculative positioning, some of which will
be transient. What it lacks is an arbitrage that ties it to today's price, so
calendar spreads, the price differences between delivery months, carry information about expected supply rather than about financing costs.

The two providers whose indices CME and ICE settle on read that shape
differently. At a panel at the FIA Expo, Silicon Data's chief executive, Carmen
Li, said the curve implied by reservation prices rises with the delivery date
for some accelerators, after falling across the board a year earlier, while
Ornn's chief technology officer, Wayne Nelms, said its data on long-term rental commitments show the hourly price falling as the contract lengthens and rising with the
size of the cluster~\citep{fia2026benchmarks}. Both compare prices for later or
longer delivery with today's price; neither compares a futures price with the
expected future rate, which is the comparison in the risk-premium argument
below. Nor is either a forward curve in the sense of this section. A schedule
of reservation prices by contract length mixes the price of later delivery with
a discount for committing to a longer term, and only futures on a fixed monthly
unit for successive months would separate the two.

The sign of the risk premium is easy to misread. Write
$F_0(T)$ for the futures price for delivery at $T$ and $E[R_T]$ for the
expected rental rate at that date. A hedger who is structurally long the good
sells forward and must pay speculators to take the other side, which pushes
the futures price below the expected future spot price,
$F_0(T) < E[R_T]$. This is Keynes and Hicks's \emph{normal backwardation}, and it
gives a long futures position a positive expected return, which is the same
statement as a positive risk premium~\citep[sec.~5.14]{hull2015}. It says
nothing about whether the futures price lies above or below today's spot rate,
which depends also on where the rate is expected to go; the terms are sometimes
used for that comparison, and it is a different one. Operators are long
capacity and want to sell it forward, so a positive premium is what hedging
pressure predicts, and the first asset-pricing treatment of compute, which
builds synthetic futures from term rental contracts, reports preliminary
evidence of one~\citep{bandisu2026}.

The premium also creates a modelling trap. A futures price is a martingale
under the risk-neutral measure, the probabilities under which every traded
asset is expected to earn the risk-free rate, and not under the real-world
measure from which historical data are drawn. A model estimated on historical
rental rates and used directly to value a forward returns the expected spot
rate and calls it the forward price, overstating the forward by the premium at
every maturity and always in the same direction. Recovering the premium needs
traded forward prices, which is one more reason the listed contracts matter.

Electricity is the closest developed analogue, sharing non-storability,
capacity-bounded supply and inelastic short-run demand, and CME places its compute contracts within its power complex.

\subsection{An accelerator-hour is not one good}

The rental rate $R^{(g)}_t$ is written as though a given accelerator model
$g$ defines a homogeneous good. It does not. Two H100-hours differ in
geographical region, interconnect fabric and topology, storage and host
specification, reliability guarantee, security posture, cluster scale, and
whether the capacity is interruptible, and buyers pay for each.

Our own collection of published rate cards shows the dispersion. Across
\HFproviders{} providers, each provider's median H100-family rate ran from
\$\HFmin{} to \$\HFmax{} per accelerator-hour, and the median of those
provider medians was \$\HFmedian{}.\footnote{Published list prices, not
transacted prices: \HFoffers{} H100-family offers in \HFdays{} daily snapshots
between \HFfirstdaylong{} and \HFlastdaylong{}, with none held between
24~September and 8~November or for 11~November. Minimum, maximum and median are
taken over one median per provider, so a provider with many regional offers
does not set the middle.}

An index must choose which of these offers it tracks, and that choice creates
\emph{basis}: the gap between the price of the exposure a hedger holds and the
price the contract settles on. (The word is unrelated to ``basis points'',
which we always write in full.) A contract settling on a neocloud on-demand
index does not hedge a hyperscaler's reserved capacity, and neither hedges
interruptible capacity. Basis risk here is not a residual to be minimised
after the fact; it is chosen when the reference index is chosen.

\subsection{The token is the better unit of account and the harder one to
standardise}

Buyers of inference want output, not accelerator-hours, and the token price
$P^{(m)}_t$ is what appears in their accounts. As a unit of economic exposure
it is the right one. It is also harder to define, because a token from a
frontier model and a token from a small model are not interchangeable: a
contract must specify capability as well as quantity, and
Section~\ref{sec:token-leg} explains why no cash-settled contract yet does.

\subsection{Why one hedge does not cover the other}

Equation~\eqref{eq:bridge} connects the two prices through throughput
efficiency $\varepsilon^{(m)}_t$ and margin $\mu^{(m)}_t$. Both move, and
neither is observable to a counterparty.

An improvement in throughput, whether from better kernels, speculative
decoding, quantisation, distillation or a change in the mix of models serving a
workload, reduces $P^{(m)}_t$ with $R^{(g)}_t$ unchanged. A capacity glut
reduces $R^{(g)}_t$ while providers hold $P^{(m)}_t$ and take the difference as
margin. In the first case a compute future does not move, so a seller of
inference who sold futures against a falling token price collects nothing. In
the second the future falls while the token price holds, so a buyer of
inference who bought futures to fix its costs loses on the hedge and saves
nothing on its tokens.

\begin{figure}[htbp]
\centering
\includegraphics[width=\textwidth]{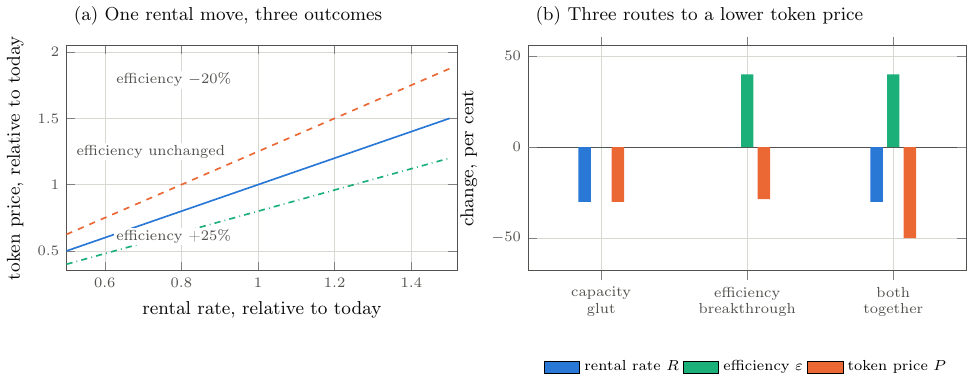}
\caption{The bridge identity, Equation~\eqref{eq:bridge}, with the margin term
$\mu^{(m)}_t$ held fixed. Panel~(a): one rental rate is consistent with many
token prices, according to what throughput efficiency $\varepsilon^{(m)}_t$
does at the same time; both axes are expressed relative to today, so the
crossing point is the present state. Panel~(b): three routes to a lower token
price, only the first of which a compute contract sees in full. Percentages
are changes from today. Both panels are computed from
Equation~\eqref{eq:bridge} rather than drawn.}
\label{fig:bridge}
\end{figure}

Figure~\ref{fig:bridge} draws both cases. The left panel holds efficiency
fixed at three levels and varies the rental rate: the compute hedge moves
along the horizontal axis, while what an inference buyer actually pays moves
along the vertical one, and the two coincide only on the middle line. The
right panel reaches a lower token price three different ways. In the middle
case the rental rate does not move at all, so a compute contract registers
nothing while the token price falls; in the third the two effects compound and
the token price falls by half on a rental move of thirty per cent.

The ratio $R^{(g)}_t / P^{(m)}_t$, which up to the margin term is throughput
efficiency itself, is thus an exposure in its own right, and no venue quotes
it.

\begin{figure}[htbp]
\centering
\includegraphics[width=\textwidth]{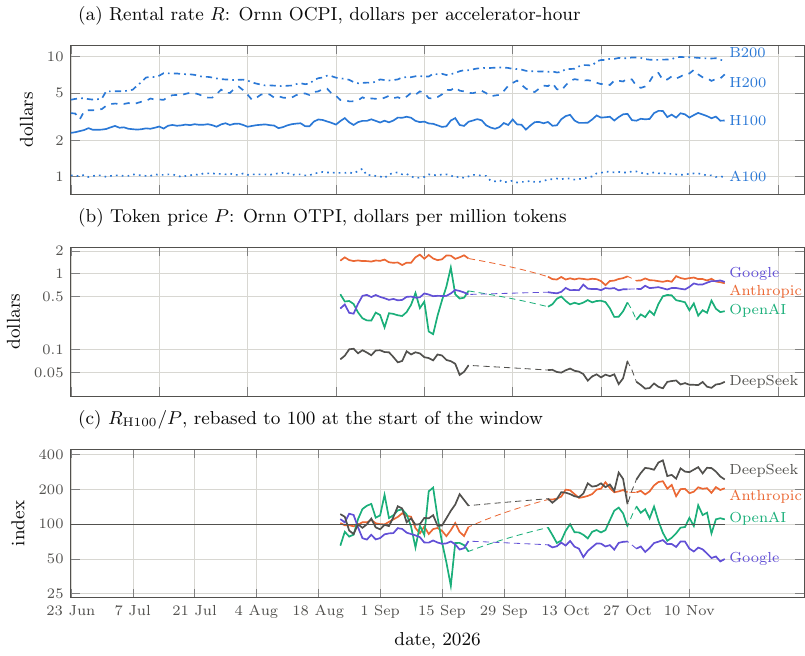}
\caption{Equation~\eqref{eq:bridge} on Ornn's public indices, keyed by the date
each publisher prints. Panel~(a) runs from \OCPIfirstdatelong{}; the token
index, and therefore panels (b) and (c), from \BSstart{}. Thin dashed segments
join the last value before and the first value after the \BSgap{} days on which
no token value is held; they are drawn for legibility only and enter no
statistic. Panel~(c) divides the H100 rate by each laboratory's token price and
rebases to 100 on the mean of its first seven observations. It equals
$\varepsilon^{(m)}_t/(1+\mu^{(m)}_t)$ only if that laboratory's tokens were
served on H100 capacity at the index rate, which is an assumption.}
\label{fig:bridge-series}
\end{figure}

Figure~\ref{fig:bridge-series} shows how the identity can be read from
published data. Ornn publishes both legs, a rental index for four accelerators
and a token price index for four laboratories, though the token index is
available to us only from \BSstart{}. The window is too short to estimate any
relationship, and we use it only to show what the identity can and cannot
reveal.

Between \BSstart{} and \BSend{} the H100 rate moved \BSdRHone\%, while token
prices moved \BSdPAnthropic\% for Anthropic, \BSdPDeepSeek\% for DeepSeek,
\BSdPOpenAI\% for OpenAI and \BSdPGoogle\% for Google, each measured between
the means of its first and last seven observations. The ratio in panel~(c) is
therefore not stable, but it cannot say why. The token index is a blend of
input, output and cached tokens across a mix of
models~\citep{techflow2026otpi}, so its level is not throughput and a move in
it can come from a list-price change, which enters $\mu^{(m)}_t$, or from a
change of mix, which moves $\mu^{(m)}_t$ and the effective $\varepsilon^{(m)}_t$
together. The pairing with an accelerator matters too: with B200 in place of
H100 the Anthropic ratio changes by \BSdQBAnthropic\% rather than
\BSdQAnthropic\%. What the figure establishes is the point of this section in
data: the two prices can move apart, and a hedge on one does not track the
other.

\subsection{Utilisation is the transmission channel}

Fleet utilisation $u^{(g)}_t$ connects the two markets to the balance sheets
financing them. An operator's revenue is the product of its installed
capacity, its utilisation and the rental rate it achieves. Debt service is
fixed. A fall in either $u^{(g)}_t$ or $R^{(g)}_t$ therefore transmits
directly to credit quality, and the two tend to fall together, because the
same excess supply that depresses the rate also leaves accelerators idle.

This is why Section~\ref{sec:deals} belongs in a document about derivatives:
the financing written against future capacity revenue determines who needs to
hedge, in what size, and when they can no longer afford to.

\section{Price formation}
\label{sec:price-formation}

\subsection{What moves the rental rate}

Five quantities drive $R^{(g)}_t$, on different timescales. \emph{Energised
capacity} sets the ceiling: it moves at the pace of grid connection rather than
of manufacturing (Section~\ref{sec:landscape}), so over any horizon a
derivative trades, supply is close to fixed. \emph{Power cost} sets the floor:
electricity is the largest variable input to operating an accelerator, so
wholesale power prices pass into rental rates, and a rental contract is in
part a power contract. The \emph{release cycle} resets the whole curve: each
new generation raises performance per unit of power and reprices its
predecessor downward, which is why the long-run level of the rental rate
declines. The depreciation debate in the accounts of the operators of
Section~\ref{sec:deals} is the same fact seen from the balance sheet, since
the useful life assumed for an accelerator is an assumption about how fast its
rental rate decays. \emph{Utilisation} converts these into realised revenue,
and \emph{contract concentration} determines how abruptly it can change: an
operator whose capacity is committed to two or three counterparties does not
experience demand as a continuous variable.

\subsection{What moves the token price}

Equation~\eqref{eq:bridge} makes the token price a function of the rental
rate, throughput efficiency and margin. The rental rate transmits as above.
Throughput efficiency improves in discrete steps as serving techniques and
model architectures change, and those steps have historically been large.
Margin is set competitively and compresses when providers with idle capacity
compete for the same workload.

The important structural point is one of sign. Efficiency improvement pushes
$P^{(m)}_t$ down while leaving $R^{(g)}_t$ untouched, so the two series can
move in opposite directions for extended periods without any inconsistency.
Correlation estimated over a window containing a large efficiency step will
therefore not survive out of sample.

\subsection{Price at a fixed capability}
\label{sec:fixed-capability}

A token price quoted for a model name is not a price for a fixed amount of
capability. A laboratory's small model of one year can outperform its large
model of the year before, so a series by model tier mixes a change in price with
a change in what the price buys. To separate them we hold capability fixed.

Epoch AI's Capabilities Index combines scores from more than fifty benchmarks
into one scale, anchored so that Claude~3.5 Sonnet scores 130 and GPT-5 scores
150~\citep{epoch2026eci}. We paired \CPscored{} of the \CProws{} rows in our table
of Anthropic, OpenAI, Google and DeepSeek list prices with the index's score for
the model, using the capture of \CPvintage{}. \CPunscored{} rows have no score,
and gated or coding-only models are set aside. For each threshold and day, the
series is the cheapest list price, blending input and output prices three to
one, among the models whose score reaches it.

\begin{figure}[htbp]
\centering
\includegraphics[width=\textwidth]{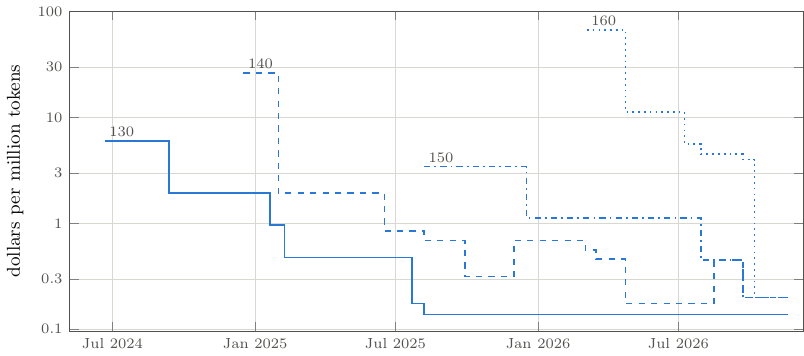}
\caption{The cheapest list price at which a given score on Epoch's
Capabilities Index can be bought, by threshold, from four laboratories' own
price lists. Prices blend input and output three to one. Each line starts on the
day the first model reaches its threshold, and steps when a cheaper qualifying
model or a price change replaces the minimum. The labels are the thresholds.}
\label{fig:capability-price}
\end{figure}

A score of 150, the level of GPT-5, was first available on \CPfirstdateC{} at
\$\CPfirstpriceC{} per million tokens (\CPfirstmodelC{}). On \CPuntil{} the
cheapest price that reaches it is \$\CPnowpriceC{}, and a year earlier the
cheapest was \$\CPagopriceC{}, a factor of \CPfactorC{} in twelve
months. The factors over the same twelve months are \CPfactorB{} at a score of 140
and \CPfactorA{} at 130.

Four limits apply. The lines are the cheapest of four laboratories' list prices,
not of the market: models served by other providers, open-weight ones among
them, are absent, so each line is an upper bound on the cheapest price
available. List prices are posted, not paid; the realised-price series we hold is
Ornn's token index, over the \BSheld{} days described in
Section~\ref{sec:compute-and-inference}. The first price at a threshold is set by
whichever model first reaches it, so a fall measured from that price overstates
the trend, and the twelve-month factor at 150 is such a fall: a year before
\CPuntil{} the cheapest price at that score was still the \$\CPfirstpriceC{} at
which \CPfirstmodelC{} first reached it.
Measured from \CPcheaperdateC{}, when a cheaper model first reached the score
at \$\CPcheaperpriceC{}, the factor is \CPcheaperfactorC{}. And Epoch
refitted its scale in November 2026, moving the scores of existing models, those
below 125 down by 13.7 points on average~\citep{epoch2026bayes}. The scores we
use postdate the refit.

These lines are a fall in $P^{(m)}_t$ at fixed capability. They do not say whether
efficiency, margin or the mix of models produced it, and they are not a fall in
$R^{(g)}_t$. The rental series we hold beside a token index covers \BSweeks{} weeks
(Figure~\ref{fig:bridge-series}) and cannot be set beside a series spanning two
years, so we make no claim about how the two moved together.
Section~\ref{sec:compute-and-inference} explains why a hedge on the rental rate
would cover such a fall only in so far as the rental rate fell with it.

\subsection{Substitution toward open weights}

A model has \emph{open weights} when its trained parameters are published, so
anyone can run it on capacity they rent or own instead of buying output from
its developer. Substitution toward such models moves demand from the token
market to the compute market without reducing it.

During 2026 that substitution became visible in usage, though not in a price
series. Over the year the share of \emph{queries} the routing service
OpenRouter sent to closed-weight models fell from around three-fifths to
roughly a quarter, and AT\&T reported open models rising from 20\% to 40\% of
its usage, with savings of up to 80\%~\citep{ftav2026openmodels}. The
benchmarking work behind that reporting puts the saving from routing between
local and cloud models at 60 to 80\% against a batched cloud
baseline~\citep{ipw2026}.

This establishes that buyers leave closed-weight models when a cheaper option
is adequate, and that the saving is large. It does not establish that the token
price has decoupled from the rental rate. Through
Equation~\eqref{eq:bridge}, the substitution compresses the laboratories'
margin $\mu^{(m)}_t$ and changes the effective throughput
$\varepsilon^{(m)}_t$ of the mix in service, both terms within the identity,
and large open models consume accelerators much as closed ones do: what moves
is who captures the margin, not necessarily how many accelerator-hours are
used. The reporting itself says the extent of the displacement ``is far from
clear'', and S\&P Global Ratings treats the threat as a known unknown rather
than a settled fact~\citep{ftav2026sandp}; we adopt that framing.

\subsection{A specification for the rental rate}

We model the logarithm of the rental rate, $X_t = \log R^{(g)}_t$, because the
rate is positive and proportional moves are the meaningful ones. Three features
must be represented. The rate reverts toward a level, because capacity is added
after high prices and withdrawn after low ones. That level declines as newer
accelerators arrive. And the rate makes occasional large moves that a diffusion
cannot produce at realistic volatility: a model launch that absorbs a fleet, a
generational transition, a regional power event.

How the third feature is modelled determines whether the model can be
estimated. The usual specification makes jumps arrive at random, with an
intensity to be estimated. But the large moves in this market are mostly
generational transitions: a new accelerator is announced and its predecessor
reprices downward, on dates known in advance. \citet{assody2026} models this
with a two-factor specification and a scheduled jump component, and reports
that the two completed transitions with public price records, the A100 across
the H100 ramp and the H100 across the Blackwell ramp, give jumps near $-0.22$
in log terms, about a fifth, against realised depreciation of the incumbent of
15 to 20\% a year. Two transitions are a very small sample, and we quote the
figure as an order of magnitude. We adopt the scheduled formulation.

Time is measured in days. Let $W_t$ be a standard Brownian motion,
$\tau_1, \tau_2, \ldots$ the announced dates of generational transitions,
$J_i$ the displacement of the log rate at transition $i$, and $\beta$ the daily
rate at which technological progress lowers the reversion target. With
$\kappa$, $\theta$ and $\sigma$ as defined in Section~\ref{sec:notation},
\begin{equation}
\label{eq:process}
\mathrm{d}X_t \;=\; \kappa\bigl(\theta - \beta t - X_t\bigr)\,\mathrm{d}t
\;+\; \sigma\,\mathrm{d}W_t \;+\; \sum_{i} J_i\,\delta(t - \tau_i)\,\mathrm{d}t .
\end{equation}
Because the reversion target falls and the rate lags it, the process settles
a distance $\beta/\kappa$ above $\theta - \beta t$ rather than at $\theta$. Where a vendor
does not pre-announce, the transition dates are unknown and arrive instead as a
Poisson process of intensity $\lambda$.

Two properties of Equation~\eqref{eq:process} matter for pricing. Because
$\kappa > 0$, the variance of $X_T$ seen from time $t$ approaches a finite
limit as $T$ grows. Implied volatility, the volatility that makes a lognormal
formula return an observed option price, has a square equal to that variance
divided by the time to maturity, so it must decline with maturity, and a
futures price far from delivery moves less than one close to it. This is the
effect \citet{samuelson1965} identified, familiar from energy markets. And
because no position in the futures reproduces a jump of unknown size, the
market is incomplete~\citep[sec.~6.9.1]{tsay2010}: no trading strategy in the
futures replicates an option payoff, and the premium for jump risk must be
estimated from option prices, of which there are none.

\subsection{Why not geometric Brownian motion}

A geometric Brownian motion has no reversion, so its variance grows in
proportion to the horizon. Here capacity responds to price, which bounds the
long-run variance. A lognormal model with one volatility applied at every
maturity would therefore overstate long-dated option values relative to
short-dated ones, and misprice calendar spreads in the same direction every
time. That does not rule out a lognormal formula at a single maturity. Without
jumps, the log rate at any one date is normally distributed under
Equation~\eqref{eq:process}, so Black's formula holds with the volatility
appropriate to that date, which is how Section~\ref{sec:products} uses it.

\subsection{Why we do not estimate it}

We state the specification and do not fit it, because the data do not yet
support estimation. Equation~\eqref{eq:process} carries four parameters,
$\kappa$, $\theta$, $\beta$ and $\sigma$, before the jump sizes and, where the
calendar is unknown, the intensity $\lambda$. Jump sizes can be identified only
from observed transitions, and the commercial rental indices span a few years
containing a handful of them. A jump size estimated from so few events has a
confidence interval spanning most of an order of magnitude, and every option
value inherits it.

Two rental indices are published free of charge, built on the opposite
principles set out in Section~\ref{sec:instruments}.
The posted-price index has \GRPIdays{} daily levels from \GRPIbasedatelong{} and
spans several accelerator families. Ornn's transaction index for the H100 has
\OCPIdays{} daily levels from \OCPIfirstdatelong{} to \OCPIlastdatelong{}.

That is enough to measure the diffusive part of the process and not the
jumps, and on the diffusive part the two series disagree by an
order of magnitude. The posted-price index returns a standard deviation of
\GRPIsigmaday{}\% a day, which annualises to \GRPIsigmaTrading{}\% on 252 trading
days or \GRPIsigmaCalendar{}\% on 365 calendar days, and to \GRPIsigmaHorizon{}\%
a year when measured over \HorizonDays-day changes. The transaction index
returns \OCPIsigmaday{}\% a day, which annualises to \OCPIsigmaCalendar{}\% on 365
calendar days, but to only \OCPIsigmaHorizon{}\% over \HorizonDays-day changes:
the variance of its \HorizonDays-day changes is \OCPIvrHorizon{} times what
independent daily moves would give, a variance ratio in the sense
of \citet{lomackinlay1988}, so most of a day's move reverses within
weeks, through fast mean reversion, noise in a fixing struck from thin
trading, or both.\footnote{Standard deviations are of log changes in the published levels; neither publisher reports a volatility. The \HorizonDays-day
figures use overlapping changes and rest on only \OCPIwindows{} non-overlapping
windows, so they are indicative. The largest single-day change of the
posted-price index is \GRPImaxmove{}\%, its publisher's own printed figure. Its
levels are the series as restated on 26~October 2026
(Section~\ref{sec:instruments}); over the 80 levels to 22~September that we
measured before the restatement, the restated index gives 7.0\% and 8.4\% a year
where the earlier construction gave 14.6\% and 17.6\%, so the construction alone
moved the measured volatility by a factor of two. Ornn's levels are those of
our latest capture, which differ from earlier captures of the same days
(Section~\ref{sec:instruments}).}

Neither series measures what an option on a one-year rental rate needs, which
is variance over a year, and at this sample length none could. What they
establish is a range. Measured volatility runs from about \GRPIsigmaHorizon{}\%
to about \OCPIsigmaCalendar{}\% a year according to how the index is built and
the horizon over which it is measured. Posted rate cards move in steps and the
index averages across families, which damps the first figure; a transaction
fixing struck from thin trading passes its noise into daily changes, which
inflates the second. The settlement convention decides which of these a contract inherits:
a contract settling on a single daily print passes the noise to its holder,
while a monthly average of daily prices, the convention in both the CME and
the ICE specification, washes most of it out. Section~\ref{sec:products} returns to the range, because the worked
example there assumes 60\% and its coupon depends steeply on that figure.

Sampling more often would not help where it matters. Intraday data would
sharpen the estimate of the diffusive volatility $\sigma$ and help to separate
continuous moves from jumps, but it cannot create generational transitions:
sampling a hundred times faster yields a hundred times the diffusive
observations and the same two launches, and the option value sits in the
discontinuous part.

We therefore carry the specification forward as a structure, not a fitted
model. A jump size or intensity calibrated to a handful of events would be
precise enough to be used and too weakly identified to be trusted.

\section{Financing raised against future capacity revenue}
\label{sec:deals}

An accelerator has almost no value apart from what it can be rented for. Its
second-hand price is set by the rental income a buyer expects to earn from it
before a newer generation arrives, so lending against accelerators is lending
against $R^{(g)}_t$. The lender is repaid in full while rental income covers
the debt and bears the shortfall when it does not, with no share in any rise:
it is in the position of someone who has sold a put on the rental rate. Tens of billions of dollars of that
exposure have been written into credit agreements since mid-2024, and the only
listed instruments on compute are offshore futures of undisclosed volume
(Section~\ref{sec:listed}). The demand for the products of
Section~\ref{sec:products} does not have to be created; it is already on
balance sheets.

\subsection{What has been raised}

Table~\ref{tab:deals} records the transactions we have been able to source. It
is a sample of representative structures, not a census. A \emph{delayed draw
term loan} (DDTL) is a committed facility drawn in tranches as assets are
delivered; an \emph{asset-backed security} (ABS) is a bond serviced by the cash
flows of a ring-fenced pool of assets rather than by its sponsor; a rating
suffixed ``(sf)'' applies to such a pool rather than to a company. Floating-rate
pricing is quoted as a spread over the secured overnight financing rate (SOFR),
the benchmark United States dollar lending rate, in basis points, each one
hundredth of a percentage point. The \DealsCompletedN{} completed transactions
total \$\DealsCompleted{} billion, of which \$\DealsTierA{} billion across
\DealsTierAN{} transactions rests on the issuers' own documents. An announced
financing plan is never added to completed transactions, and each subtotal is a
floor on its category.

\begin{table}[htbp]
\centering
\footnotesize
\setlength{\tabcolsep}{5pt}
\caption{Financing raised against compute capacity and the facilities housing it. Face amounts are committed, not drawn. The source column says whether the face amount comes from the issuer's own filing or announcement or from a press report. An announced financing plan is not a completed issue and is subtotalled separately; one that mixes debt with equity is shown with the collateral of its debt.}
\label{tab:deals}
\begin{tabular}{@{}lllrlll@{}}
\toprule
Announced & Sponsor & Instrument & \$bn & Pricing, rating & Collateral & Source \\
\midrule
2024-05 & CoreWeave & DDTL & 7.50 & SOFR+600/650;+1300bp & unspecified & issuer \\
2025-05 & CoreWeave & unsecured notes & 2.00 & 9.250\% fixed & unsecured & issuer \\
2025-07 & CoreWeave & DDTL & 2.60 & SOFR+400bp & all assets & issuer \\
2025-10 & Meta & secured notes & 27.30 & 6.581\% fixed, A+ & real estate & press \\
2026-02 & Oracle & financing plan & 50.00 & IG & unsecured & issuer \\
2026-03 & CoreWeave & DDTL & 8.50 & SOFR+225bp, A3 & accelerators & issuer \\
2026-04 & Switch & ABS & 0.77 & AAA (sf) & real estate & issuer \\
2026-04 & Core Scientific & secured notes & 3.30 & 7.750\% fixed & unspecified & issuer \\
2026-06 & Applied Digital & secured notes & 1.59 & 7.000\% fixed & unspecified & issuer \\
2026-07 & Nebius & secured facility & 0.78 & SOFR+250bp & accelerators & issuer \\
2026-07 & Meta & secured notes & 12.50 & 7.534\% fixed, A+ & real estate & issuer \\
2026-08 & CoreWeave & DDTL & 2.60 & Term SOFR+550bp, Ba2 & all assets & issuer \\
2026-08 & Nebius & convertibles & 5.75 & 0.50\%/4.50\% & unsecured & issuer \\
2026-09 & CoreWeave & convertibles & 3.70 & 2.875\% fixed & unsecured & issuer \\
\midrule
\multicolumn{3}{@{}l}{Completed transactions} & 78.88 & \multicolumn{3}{l}{13 transactions} \\
\multicolumn{3}{@{}l}{\quad of which from issuers' own documents} & 51.58 & \multicolumn{3}{l}{} \\
\multicolumn{3}{@{}l}{Announced plan, not yet issued} & 50.00 & \multicolumn{3}{l}{excluded from the above} \\
\bottomrule
\end{tabular}
\end{table}

\begin{figure}[htbp]
\centering
\includegraphics[width=\textwidth]{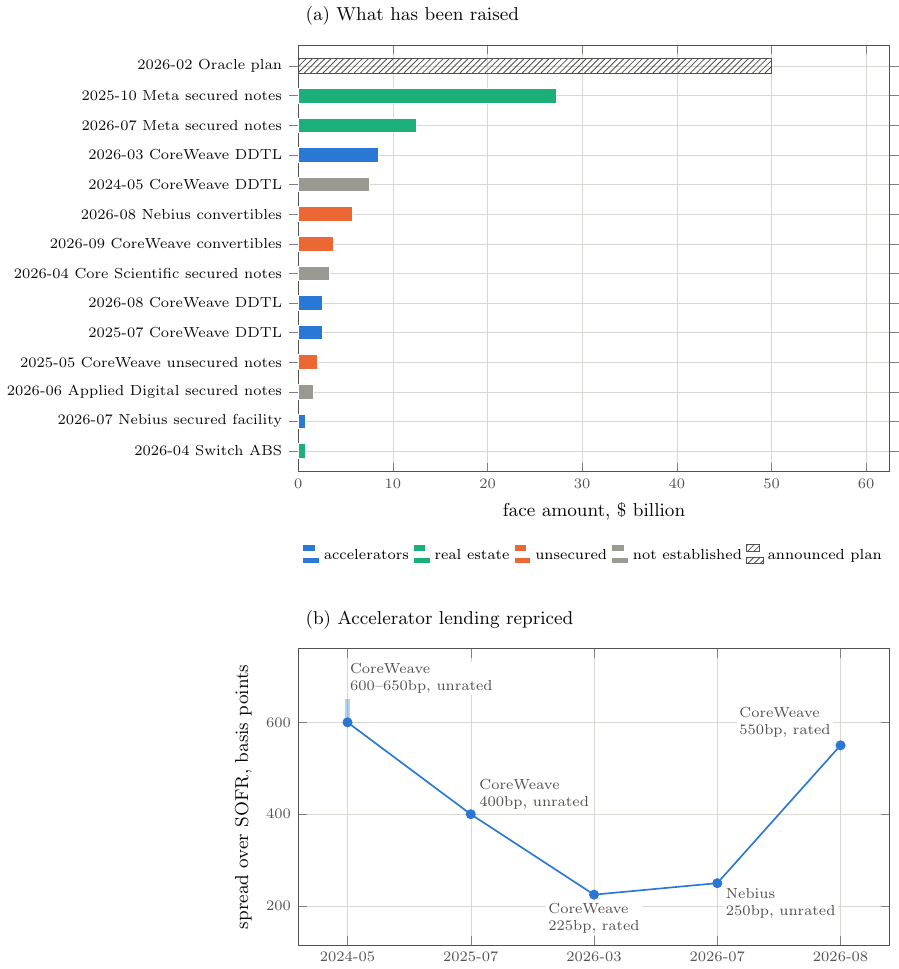}
\caption{The transactions of Table~\ref{tab:deals}, built from the same
records. Panel~(a): face amount by transaction, coloured by what secures it;
grey marks collateral we could not establish from a primary source, which is
not the same as unsecured, and the hatched Oracle bar is an announced plan.
Panel~(b): the spread over SOFR for the facilities that record one, labelled by
sponsor and by whether an agency had rated the facility. The May 2024 facility
priced off a grid keyed to the credit of the contract each draw serves; the
shaded bar spans its investment-grade tier, and its non-investment-grade tier,
at 1{,}300 basis points, is off the scale.}
\label{fig:deals}
\end{figure}

Five structures appear. \emph{Accelerator-secured term loans} lend against the
hardware and the contract it serves, to a borrower that is
\emph{bankruptcy-remote} (structured so the lender's claim survives the
sponsor's insolvency) and usually \emph{non-recourse} (enforceable against the
vehicle's assets and not the parent); CoreWeave's successive DDTLs are the
archetype, and Nebius's first secured facility is the same idea with the parent
as borrower~\citep{nebius2026secured}. \emph{Corporate unsecured issuance} funds
capacity from the sponsor's own credit, as in CoreWeave's 9.250\% senior notes
due 2030, the convertible issues of Nebius and CoreWeave, and the senior
unsecured bonds in Oracle's announced plan to raise \$45 to \$50 billion in
calendar 2026, about half of it in
equity~\citep{crwv2025notes,crwv2026conv,oracle2026plan}. \emph{Project notes}
finance one facility at the entity that owns it, with the parent guaranteeing
completion rather than repayment, as in Applied Digital's \$1.59 billion issue
against a North Dakota building~\citep{apld2026notes}. \emph{Property
securitisation} finances buildings rather than machines: Switch's Series 2026-1
raised \$768 million against eleven data centres serving more than 550
customers, roughly 84\% of trust revenue from investment-grade tenants, with a
AAA(sf) senior tranche~\citep{switch2026abs}. And \emph{off-balance-sheet joint
ventures} move the largest projects off the sponsor's accounts, as in Meta's
Hyperion arrangement with funds managed by Blue Owl
Capital~\citep{meta2025hyperion}.

The accelerator rows and the property securitisation are both called secured
financing of AI infrastructure, but they are not the same risk. The
securitisation rests on real property and a diversified, mostly
investment-grade tenant base; the accelerator facilities rest on one offtaker
and one depreciating asset class.

\subsection{The asset class repriced, and then split}

The clearest signal in the table is how far one sponsor's secured borrowing
repriced while its structure stayed the same: non-recourse loans to a
special-purpose vehicle, drawn as accelerators are delivered and serviced by
the contracts they run. CoreWeave's May 2024 facility of \$7.5 billion priced
off a grid, at SOFR plus 600 to 650 basis points where the contract a draw
served was investment grade and 1{,}300 where it was not, and was
unrated~\citep{crwv2024ddtl2,crwv2025ddtlamend}. Its July 2025 facility of
\$2.6 billion priced at SOFR plus 400, unrated~\citep{coreweave2025ddtl3}. Its
March 2026 facility of \$8.5 billion priced at SOFR plus 225, about 5.9\% fixed,
rated A3 by Moody's and A~(low) by DBRS~\citep{coreweave2026ddtl}, and Nebius's
facility of July 2026 priced within 25 basis points of it. That is 175 basis
points of compression in eight months and about 400 over twenty-two, as
accelerator-backed paper went from unrated to single-A. Nothing in the
underlying, whether useful life, rental-rate volatility or offtaker
concentration, changed in anything like that proportion: the compression is a
repricing of perception, and perception reprices in both directions.

It did. An incremental \$3.0 billion tranche added to the May 2024 agreement in
September 2025 priced at SOFR plus 425, wider than the facility two months
before it~\citep{crwv2025ddtlamend}. And in August 2026 CoreWeave's DDTL~5.5
raised \$2.6 billion at SOFR plus 550, rated Ba2 by Moody's and BB+ by Fitch,
325 basis points wider than the March facility and below investment
grade~\citep{crwv2026ddtl55}. By the issuer's account the earlier facilities
were backed by customer contracts running to the maturity of the debt, while
this one runs about five years against contracts averaging about three
(Section~\ref{sec:ddtl}). Sponsor and structure are the same; what differs is
whether the offtake covers the loan. An investment-grade rating on such a
facility is a statement about the offtaker, not about the hardware, and when
the contracts end before the debt the rating falls from single-A to double-B.

The compression was not market-wide either. Meta financed two campuses
through the same structure, each rated in the A category: Hyperion in October
2025 at 225 basis points over Treasuries, and El Paso in July 2026 at
287.5~\citep{ifr2025beignet,meta2026elpaso}.\footnote{Both spreads come from
trade press; Meta's own release states the size and structure of the El Paso
financing but not its pricing.} Over a period overlapping CoreWeave's
compression, the same sponsor's property debt widened.

\subsection{What the loan documents price, and what trades on it}

The rows of Table~\ref{tab:deals} rank differently in a capital structure.
Safest is secured debt at the vehicle that owns the assets, CoreWeave's DDTLs
and Applied Digital's project notes; behind it comes senior unsecured debt of
the operating company, with convertible notes ranking alongside; and equity
takes the first loss. One loan document states the dependence
on the rental economy outright. The pricing grid of CoreWeave's May 2024
facility is keyed not to the borrower's credit but to the credit of the
contract each draw serves, and prices a non-investment-grade offtaker at
roughly double the spread~\citep{crwv2025ddtlamend}: the lender wrote the
offtaker's credit into the interest rate.

The same sponsor sold secured and unsecured debt two months apart, which gives
a rough price for the collateral. Converting its 9.250\% unsecured notes of May
2025 to a floating spread with the 3.65\% reference rate implied by its March
2026 facility, quoted both ways, puts them near SOFR plus 560, against SOFR
plus 400 on the secured facility of July 2025: about 160 basis points for the
security interest~\citep{crwv2025notes,coreweave2025ddtl3,coreweave2026ddtl}.
The conversion anchor is ten months later, the maturities differ, and the
unsecured notes are a claim on the parent rather than the vehicle, so the
figure is a magnitude rather than an estimate.

A desk with a view on capacity prices can express it in credit today, but the
instruments sit in the wrong names. Single-name credit default swaps, contracts
that pay out if a named issuer defaults, now exist on several of these issuers,
CoreWeave's from late September 2025~\citep{ifr2026cds}. Oracle's is the liquid
one: in March 2026 it was the most actively traded single name among United
States corporates, at 311 trades against 138 for the next~\citep{clarus2026cds},
and protection outstanding on it runs to tens of billions of dollars against
hundreds of millions on CoreWeave~\citep{ifr2026cds}. No single-name contract
could be established on Nebius, which sells accelerator capacity directly, or
on Applied Digital, the most dependent name in the screen of
Section~\ref{sec:landscape}. The standard North American
investment-grade index holds Oracle among 125 names chosen for liquidity, so
protection bought on it is mostly protection on unrelated
credit~\citep{ice2026cdxig}. Single-name protection is thin in general, with
fewer than three per cent of reference entities averaging more than ten trades
a day over the five years to 2023~\citep{iosco2025cds}, and protection bought
from a counterparty exposed to the same downturn carries \emph{wrong-way risk}:
the protection seller is likely to be under strain in exactly the state in
which the protection pays.

There is, however, traded volatility on these issuers, and it arrived through
convertible bonds, which are debt with a share option attached. CoreWeave's
September 2026 issue of \$3.7 billion of 2.875\% notes due 2033 carried a
conversion premium of about 22.5\% and capped calls costing about
\$499 million, roughly 13.5\% of the principal, and the release records that
dealers expected to hedge in the shares~\citep{crwv2026conv}. It is volatility
on the equity rather than on the rental rate, but it is the only option
position in this complex that can be put on at size, which is why
Section~\ref{sec:products} returns to it.

We found no work relating rental rates to the credit spreads of the operators
borrowing against them; the one asset-pricing treatment of compute does not
examine credit~\citep{bandisu2026}. Both series exist, and the
variance-minimising hedge ratio of Section~\ref{sec:products} is computable on
them today.

\subsection{Circularity}

Sona Asset Management catalogued 176 transactions totalling \$3.6 trillion
between 202 entities over the three and a half years to August 2026, about 120
of them flagged as highly circular, with capital flowing in a loop between
counterparties connected several ways~\citep{ftav2026sona}. Amazon invested
about \$50 billion in OpenAI while securing a compute commitment of about
\$100 billion from it; Microsoft's recapitalisation of OpenAI, about
\$135 billion, sits beside an OpenAI commitment of about \$250 billion to Azure;
and NVIDIA committed conditionally to invest up to about \$100 billion in
OpenAI, which in turn committed to Oracle, Microsoft and Amazon, among the
largest buyers of NVIDIA's chips. These are equity and purchase commitments,
not debt raised against capacity, and they stay out of Table~\ref{tab:deals}.
Circularity is not fraud, but it means no node's credit can be assessed in
isolation, and a proxy hedge chosen from within this web is not independent of
the exposure it is meant to offset.

\subsection{What the rating agencies now say}

S\&P Global Ratings has published downgrade thresholds for the six largest
United States hyperscalers, in adjusted debt to EBITDA: Oracle may reach 4.5
times, SpaceX 2 times, Amazon 1.5 times and the rest 1
time~\citep{ftav2026sandp}. Oracle is the consequential case, because a
downgrade from BBB$-$ would take roughly \$117 billion of its bonds out of
investment-grade indices and force index-tracking funds to sell. The agency also states why operators are natural buyers of protection: customer
contracts run three to five years against a data-centre life of twenty or
more, so the asset is financed over a horizon no customer has contracted.

We infer, as a hypothesis and not a finding, a channel from the rental rate to
forced selling: lower rental revenue compresses EBITDA, which approaches a
published threshold, a breach of which would oblige index trackers to sell.
The agency's own framing is that thresholds are approached through rising
commitments or an earnings miss generally, not specifically through rental
rates, and a threshold is not a trigger. The channel bites on leveraged
operators, whose debt service is fixed against floating revenue, and not on
hyperscalers funding fleets from operating cash flow; Oracle appears on the
list because of its leverage, not because its rental income is unusually
exposed.

\section{Demand and supply for risk transfer}
\label{sec:risk-transfer}

The preceding sections describe a physical market and the debt raised against
it. Whether a derivatives business exists there depends on a different
question: who is structurally long or short the rental rate, in what size, over
what horizon, and with what capacity to bear it. A large physical market can
carry no hedging demand if every participant is naturally flat, and a small one
can carry a great deal if they all sit on the same side.

We establish three things. Hedging demand is a duration mismatch, concentrated
in operators whose debt runs past their disclosed contracted revenue; the one
single-year shortfall that filings alone prove is a refinancing need rather
than debt outliving its contracts. The mismatch
cannot be measured from published information, because the two legs
are disclosed under different accounting standards. And in one publicly
syndicated loan, lenders finance the uncontracted tail and settle it by cash
prepayment against a private model, because the rental rate, unlike the
interest rate and the energy price beside it, has no market to hedge it in.
We do not establish demand: exposure is not willingness to pay. The evidence
covers listed operators only, while our price collection shows private
providers to be the majority by count, and no filing we read states the price
mechanism in its main contracts, so we quote no aggregate figure for the
hedging demand $D$ defined below.

\subsection{Hedging demand is a duration mismatch, not a fleet size}

An operator that has sold capacity forward at a fixed price under a
multi-year contract has hedged its price risk for the life of that contract --- contractually, not
financially. Its residual exposure is capacity not under contract, and the
years by which its debt outlives its contracted revenue. Hedging demand is
therefore not proportional to installed fleet. It is

\begin{equation}
\label{eq:hedgingdemand}
D \;=\; \sum_{\text{operators}} \; \text{financed capacity} \;\times\;
\bigl(\text{years of financing} - \text{years of contracted offtake}\bigr)^{+} ,
\end{equation}

where $(\cdot)^{+}$ denotes the positive part, because an operator whose
contracts outlast its debt has no exposure to measure.
Section~\ref{sec:deals} already records the agency observation that makes this
quantity large: customer contracts run three to five years against
data-centre lives of twenty or more, with the financings in between.

\subsection{The two legs are disclosed under different accounting standards}

Equation~\eqref{eq:hedgingdemand} looks computable from public filings. It is
mostly not, for a structural reason.

Debt maturity disclosure is standardised and granular: every filer publishes
five years plus a residual. Contracted revenue disclosure is not standardised
at all. Across the six operators we examined we found four different measures
in use --- remaining performance obligations under the revenue standard, lessor
lease receivables under the lease standard, maximum consideration under named
contracts, and megawatts or a single transaction's press-release value. The two
legs of the mismatch are therefore measured on different bases at different
granularities, and only one company in the set, Applied Digital, publishes a year-by-year
run-off of its contracted revenue on the same grid as its debt table.

The asymmetry shows in the tagging itself. CoreWeave reports its debt
maturities as a machine-readable ladder --- the remainder of this fiscal year,
each of the next four, and everything beyond --- and reports its contracted
revenue as a single undated total. One leg arrives as a schedule; the other
arrives as a number.

That also gives us a way to derive the undated bucket rather than read it off a
table. At \CWdebtdate{} the issuer tags total debt of \$\CWdebttotal{}~billion
against dated maturities summing to \$\CWdebtdated{}~billion, leaving
\$\CWdebtresidual{}~billion with no disclosed year: \CWdebtpct\% of the total,
which is the CoreWeave row of Table~\ref{tab:undated}.

The consequence is commercial, not only a matter of data: \emph{the market
cannot see its own duration mismatch}, and neither can its lenders on
published information alone.

\subsection{The one shortfall the filings prove}

Applied Digital is the only operator in our sample whose debt schedule and contracted revenue
schedule share a fiscal-year grid, so no interpolation is
required~\citep{apld2026tenk}. In its fiscal year 2031, ending 31~May, scheduled debt principal of \$4.19~billion exceeds contracted lease receipts of
\$2.30~billion by about \$1.9~billion (Table~\ref{tab:apldmismatch}, in \$~thousand).

The cause is bullet maturities bunching: two note issues and a convertible fall
due in one fiscal year. Its leases run past its debt, so
Equation~\eqref{eq:hedgingdemand} counts nothing for it; the shortfall is a
refinancing need, whose cost depends on the credit of its tenants, CoreWeave
among them, and through it on the rental rate. Adding the \$1.59 billion issued in June
2026~\citep{apld2026notes}, which also matures in 2031, widens the shortfall to
approximately \$3.5 billion. The table is struck at 31~May and so omits a
fifteen-year, 210-megawatt lease of approximately \$5.2~billion signed on
5~June~2026~\citep{apld2026tenk}; spread evenly over its base term, that
narrows the fiscal-2031 gap by some \$0.35~billion, to approximately
\$3.1~billion.

Core Scientific, a colocation lessor outside that sample, shows why a shared
grid is not enough. It reports both legs by calendar year, but its schedule of
lease receipts ends at 2030, and \$3.93~billion of its \$4.39~billion of notes
falls due in 2031, the first year the schedule does not date: \$3.30~billion
of secured notes in May and \$0.63~billion of convertible
notes~\citep{corz2026tenq}. Against that maturity the schedule offers only
\$4.24~billion of receipts spread across every later year, before a further
\$3.1~billion from twelve-year leases not yet commenced. The mismatch is dated
on the debt side and undated on the other.

\subsection{The larger exposure is the part nobody dates}

{\widowpenalty=10000 \clubpenalty=10000
The provable gap is not the largest one. Across the sample, a substantial share
of debt matures in an undated residual bucket, while the revenue that would
service it is dated, if at all, only in coarse bands.\par}

\begin{table}[t]
\centering
\footnotesize
\caption{Applied Digital: scheduled debt maturities against contracted lease
receipts. The FY2031 shortfall is the only single-year shortfall provable from
filings in our sample; the leases outlast the debt, so it is a refinancing
need. Gross debt principal against gross lease receivables;
refinancing and new contracts are ignored.}
\label{tab:apldmismatch}
\begin{tabular}{@{}lrrrrrr@{}}
\toprule
 & FY2027 & FY2028 & FY2029 & FY2030 & FY2031 & Thereafter \\
\midrule
Debt maturing & 16{,}591 & 196{,}648 & 290{,}798 & 295{,}689 & 4{,}194{,}954 & 312{,}000 \\
Lease receipts & 451{,}065 & 1{,}453{,}748 & 2{,}251{,}769 & 2{,}280{,}829 & 2{,}299{,}574 & 27{,}107{,}897 \\
\midrule
Shortfall & --- & --- & --- & --- & 1{,}895{,}380 & --- \\
\bottomrule
\end{tabular}

\vspace{1.2em}
\caption{Debt maturing beyond the disclosed horizon. Percentages are of each
company's total debt principal at its most recent balance sheet date, computed
from each issuer's own tagging as total debt less the buckets it dates. A dash
means the two cannot be put on one basis: TeraWulf's most recent tagged
carrying amount predates its dated buckets, and we leave the row empty rather
than mix reporting dates. How far each company's own contracted-revenue
disclosure extends is not computable from the tagging and is given in the text
below.}
\label{tab:undated}
\begin{tabular}{@{}lrr@{}}
\toprule
 & Residual bucket & Share of debt \\
\midrule
IREN & \$6.43bn & 64\% \\
CoreWeave & \$19.64bn & 41\% \\
TeraWulf & --- & --- \\
Nebius & not disclosed & --- \\
\bottomrule
\end{tabular}

\end{table}

What the tagging cannot show is how far each company's own disclosure of
contracted-revenue timing extends. IREN
gives none at all for its \$11.4 billion lease component. CoreWeave's runs to
seventy-eight months but in three bands, the last of which places a fifth of
its \$103.7~billion backlog somewhere in months 49 to 78~\citep{crwv2026tenq}. TeraWulf
discloses no portfolio figure, and Nebius no annual schedule. The debt side is a schedule in every case, and the revenue side is a
year-by-year schedule in none.
The largest backlog in the sector is dated no better: Alphabet reports
\$519.5~billion of remaining performance obligations at 30~June~2026,
\$513.9~billion of it Google Cloud, and says only that just over half falls
within twenty-four months~\citep{goog2026tenq}.

IREN is structurally the most exposed: most of its debt has no matching revenue
timing at all: at \IRENdebtdate{}, \IRENdebtpct\% of \$\IRENdebttotal{}~billion fell beyond the
years it dates, against a debt base that grew roughly
eightfold in one fiscal year, from \$0.99 billion to \$7.71
billion~\citep{iren2026tenk}, and by a further 30\% in the following quarter.

\subsection{How long contracts actually run}

Contract duration varies widely by facility type. Weighted-average remaining contract terms disclosed in
rating actions run from 2.0 years for a multi-tenant retail colocation
portfolio and 2.4 years for another, through 3.9 years for a third, to
approximately 13.8 years for a single-tenant hyperscale
property.\footnote{Rating agency rating-action releases
\citep{kbra2026databank,kbra2026flexential,kbra2026centersquare,kbra2026lohrasp}.}
An industry primer gives the same spectrum from the pricing side: hyperscale
capacity at \$100 to \$150 per kilowatt-month on ten to fifteen year triple-net
terms, wholesale colocation at \$150 to \$250 on five to ten years, and retail
colocation at \$200 to \$400 on one to three years, with base rent
supplying 70 to 80\% of hyperscale revenue and metered power passed through
separately.\footnote{An industry association primer~\citep{crefc2026primer}.}

Two things follow. The mismatch of Equation~\eqref{eq:hedgingdemand} is
structurally worst at the short-tenor end, where a two-year contracted book
supports five- to seven-year debt. But the short-tenor end rents space and
power rather than accelerators, so its exposure is to data-centre rent and not
to $R^{(g)}_t$. It is the same structure of risk on a different underlying, and a candidate
for a separate product.

\subsection{What operators already pay for certainty}

The best available evidence on the price of revenue certainty comes from an
operator disclosing its own contract economics. Nebius reports long-term
committed capacity at \$20 to \$25 million per megawatt of annual contract
value against \$40 to \$50 million per megawatt for short-term capacity, and
states that roughly 70\% of one quarter's deals carried customer prepayments
covering 50 to 60\% of the associated capital expenditure, with more than \$9
billion of prepayments expected across the year, in its shareholder letter
rather than its financial statements~\citep{nbis2026letter}.

That is a steep term structure of physical capacity prices: the long-term
rate is roughly half the short-term one. The disclosure does not say which side
pays. On one reading, the operator accepts a lower rate in
exchange for revenue certainty and is already paying something of that size for
it. On the other, the buyer of immediate capacity is paying a scarcity premium
and the operator earns it by holding capacity back, spending nothing. Some of
the gap is neither, being scale, credit quality and service level rather than
term at all, and we do not decompose it.

Under the first reading a financial hedge offers an operator a different way
to buy what it already buys, and the physical discount bounds what the hedge
can charge. Under the second, the hedge has to beat simply withholding
capacity. The prepayment figure makes a further point: a prepaid forward on
capacity would not have to be introduced to this market, since it is already
among its main financing forms.

\subsection{Take-or-pay fixes volume, not price}
\label{sec:mechanism}

Operators rarely say whether the price in a customer contract is fixed for the
term, subject to escalators, or reset against an index. We read the most recent
annual or quarterly filing of four listed operators for the answer. None gives it for
its main contracted book.

Applied Digital reports approximately 1{,}410 megawatts of contracted critical
IT load across five campuses and approximately \$36.2 billion of total
contracted revenue over initial fifteen-year base lease terms, every lease held
directly with CoreWeave or an investment-grade hyperscaler, each structured on
a take-or-pay basis and non-cancellable, so that a termination for convenience
requires payment of the full remaining contractual value~\citep{apld2026tenk}.
Its revenue policy states a fixed rate only for two smaller lines, co-hosting
and the managed GPU cloud of its consolidated affiliate ChronoScale. For the
leases that make up the contracted book it states no price mechanism.

CoreWeave uses the same term, likewise without the price. It sells capacity ``through
committed contracts, which are take-or-pay, or on-demand, which are
pay-as-you-go'', and committed contracts supplied 98\% of its revenue in the
second quarter of both 2025 and 2026~\citep{crwv2026tenq}. The filing does not
say how the rate in those contracts is set. IREN does not use the term at all:
its reserved contracts ``generally specify'' capacity, service levels, pricing,
term and prepayments, and again the mechanism is not
stated~\citep{iren2026tenk}. TeraWulf, outside the four, states leases of ten
to twenty-five years with contractual escalators but gives no
rate~\citep{wulf2025tenk}. The smallest operator we read, Axe
Compute, describes ``take-or-pay provisions intended to provide income
visibility'' and is the only one to schedule its remaining performance
obligations year by year, \$60.8~million of them, again with no price
mechanism stated~\citep{agpu2026tenq}.

Nebius discloses an arrangement that transfers utilisation rather than price. Under an agreement of 13~March~2026, Meta took dedicated GPU clusters, with associated
storage and connectivity, for five years at a total contract value of \$12~billion, and under a
further order undertook that where capacity in certain clusters ``is not sold
by Nebius to other customers, Meta is obligated to purchase such unsold
capacity'' for the remainder of five years from its deployment, a potential
contract value of up to \$15~billion~\citep{nbis2026q1}. Meta has in effect
written Nebius a five-year put on the capacity it cannot sell, at a price the
filing does not state. It is one of two such arrangements we can document from
a statement of record; in the other, NVIDIA, CoreWeave's supplier and a
stockholder, is obliged to buy CoreWeave's residual unsold capacity through
13~April 2032, at an initial value of \$6.3~billion~\citep{crwv2025nvidia}.
Both hedge quantity, not the rate.

Take-or-pay is therefore general but not universal, and where it appears it
contracts away volume risk rather than price risk. If a book is also fixed-price, neither party holds a residual rate exposure
within the term, the exposure sits at renewal, and
Equation~\eqref{eq:hedgingdemand} measures all of it. If prices reset within
the term, exposure sits inside the contracts as well and the equation is a
lower bound. No filing we read says whether any book is, and the
silence is probably structural. Meta's subsequent-event note on the Hyperion joint
venture states the initial lease term, the renewal ceiling, the commitment
amount and a sixteen-year residual value guarantee with its opening threshold,
and says nothing about escalation or indexation~\citep{meta2025tenq}.
A filer that discloses this much about its largest such arrangement does not
omit a pricing mechanism by oversight; the mechanism is not a required
disclosure. A figure of 3\% circulates in secondary
commentary and is attributed to that filing. It does not appear in it, and we
do not use it.

\subsection{The one renewal risk with a price}
\label{sec:ddtl}

One financing now prices the positive part in
Equation~\eqref{eq:hedgingdemand} directly. On 7~August~2026 a CoreWeave
subsidiary entered a \$2.6~billion delayed-draw term loan maturing on
1~September~2031, at Term SOFR plus 5.50\%, guaranteed by the parent and
secured on the borrower's assets~\citep{crwv2026ddtl55}. The issuer's
announcement states what distinguishes it: earlier facilities of this kind were
backed by customer contracts running to the maturity of the debt, whereas this
one runs approximately five years against customer contracts averaging
approximately three, and in financing the difference the lenders show ``a
willingness to underwrite renewal risk''. The tenor comparison comes from the
accompanying press release; the loan terms come from the credit agreement. On the issuer's own averages, some two of the
facility's five years are financed capacity with no contracted offtake behind
it, and a facility arranged by JPMorgan and MUFG finances that tail at
550 basis points over the reference rate for the loan as a whole.

The credit agreement shows how the lenders protect themselves against the tail,
and the protection is not a hedge. If a customer contract expires or terminates and is not renewed or
replaced, by a deadline itself redacted, with a contract that keeps a projected
debt service coverage ratio of at least 1.35 and a ratio of projected contract
value to outstanding loans of at least 2.40, the borrower must prepay the loan
within five business days by whatever amount restores both. A replacement counts only
if it runs at least six months and is on terms ``not materially worse, taken as
a whole'' than the contract it replaces. For uncontracted periods the projected
ratio is the borrower's good-faith determination on assumptions ``reasonably
consistent with the Base Case Model'', a model delivered to the agent and lenders at closing and not filed. Renewal risk is thus settled in cash against a forward
rental rate that no market quotes and only the parties see.

The agreement handles the facility's other price exposures differently. The
borrower must hold interest-rate hedges on at least 95\% of projected principal
within forty-five days of the end of the drawing period, and it may hedge ``the price of energy or capacity'' with its lenders or
other suitably rated counterparties, secured on the same collateral. No
provision addresses the rental rate expressly, although that permission does
not exclude one. The agreement defines a ``Weighted Average Hourly GPU Rate''
but redacts the definition, and the term appears nowhere else in the filed text. It
also redacts the ``Total Projected Contract Value'' on which the 2.40 test
rests, and the counterparties to all five initial customer contracts. IREN's
filing shows the same pattern, with some \$1.5~billion of pay-fixed
interest-rate swaps to October 2031 designated as cash-flow hedges and no
instrument on the rental rate~\citep{iren2026tenk}. The legs a market exists to hedge are hedged; the one
without a market is handed back to the borrower as a prepayment obligation.

The issuer adds one datum in passing: shorter customer contracts ``often
command higher prices'', which agrees in sign with the Nebius term structure
above without settling which side pays for it.

\subsection{What becomes observable once contracts trade}

One measurement arrives on its own. Positions in a listed futures contract are
reported to the regulator above a threshold, and aggregated weekly into a public
report splitting open interest between commercial hedgers and other categories.
The default reportable level for a commodity without a market-specific level is
25 contracts~\citep{cfr1503}, and the report is published weekly on the third
business day after the Tuesday it describes~\citep{cftccot}.

That is a direct observation of exactly what this section has had to infer from
filings: who is hedging, on which side, in what size. It will be thin at first
and will not qualify for publication until enough reportable traders exist. It is nonetheless the most valuable series the market will produce, and it
should be captured from the first week of trading, because early weeks cannot
be recovered later.

\section{A note that sells protection on the rental rate}
\label{sec:products}

Section~\ref{sec:deals} showed that a large exposure to the rental rate sits in
credit agreements, held by operators who cannot lay it off, and
Section~\ref{sec:ddtl} showed one such agreement settling that exposure by cash
prepayment rather than by a hedge. Section~\ref{sec:instruments} showed that,
once the CME and ICE contracts trade, they will offer a linear hedge on two
accelerator models and no option. This section sets out an instrument
through which an operator could buy the protection it lacks, and what it would
cost.

\subsection{The instrument}

A reverse convertible note pays an enhanced coupon and returns principal in
full unless the reference price finishes below a strike, in which case the
holder absorbs the shortfall. It is a bond plus a plain vanilla put sold by
the holder, not an exotic option~\citep[ch.~15]{deweert2008}.

Write $r$ for the continuously compounded risk-free rate, $T$ for the term of
the note in years, and $F_0(T)$ for the forward rental rate for settlement at
$T$. Principal $N$, coupon rate $c$ and strike $K$ are as defined in
Section~\ref{sec:notation}. Throughout this section a single accelerator model
is understood, so we write $R_T$ for $R^{(g)}_{T}$ and $F_0(T)$ for
$F^{(g)}_{0}(T)$. The holder's redemption amount is
\begin{equation}
\label{eq:rcpayoff}
N \min\!\left(1, \frac{R_T}{K}\right)
\;=\; N \;-\; \frac{N}{K}\,\max\!\left(K - R_T,\, 0\right).
\end{equation}
The right-hand side is the identity that matters: the note is a zero-coupon
bond of face $N$, less $N/K$ European puts struck at $K$, plus the coupon. The
put notional is the principal divided by the strike, so that exercise exactly
offsets the bond~\citep[sec.~15.1]{deweert2008}. The holder is short the puts
and the issuer is long them. We let $R_T$ be, like the settlement of the listed
futures, the average of the daily index over the note's final month, so that
those futures hedge it without a difference in fixing. Averaging over the last
month of a one-year term lowers the effective volatility by about three per
cent of itself, which we ignore, so the valuation below slightly overstates the
put.

Let $p(K,T)$ be the value today of one such put. Issuing the note at
\emph{par}, 100 per cent of its principal, requires
\begin{equation}
\label{eq:coupon}
N \;=\; N e^{-rT} \;+\; c\,N\,A(T) \;-\; \frac{N}{K}\,p(K,T),
\end{equation}
where $A(T)$ discounts the coupon stream and equals $e^{-rT}$ for a single
terminal coupon. Solving for $c$ separates the coupon into two parts:
\begin{equation}
\label{eq:couponsolve}
c \;=\; \underbrace{\frac{1 - e^{-rT}}{A(T)}}_{\text{time value of money}}
\;+\; \underbrace{\frac{p(K,T)}{K\,A(T)}}_{\text{put premium earned}} .
\end{equation}
The coupon above the risk-free rate is not a reward for holding the note; it is
the price the issuer pays for the protection the holder has sold. It rises with
the rate of interest and with the put premium, and so with volatility and with
a strike closer to spot~\citep[sec.~15.3]{deweert2008}. No configuration makes
the enhancement free.

\subsection{A worked example}

Take $N = \$1{,}000{,}000$, a one-year term, spot and one-year forward rental
rates both at a round \$3.20 per accelerator-hour, close to \$\HFmedian{}, the
median of the per-provider H100-family medians in our collection, a strike at
80\% of spot so that $K = \$2.56$, a risk-free rate of 4\%, and a volatility of
60\% a year. We value the put with Black's model for an option on a
forward~\citep[sec.~18.8]{hull2015}, which assumes the rate is lognormal at
expiry. The volatility it takes is the one-year figure, the annualised standard
deviation of the log rate a year ahead, which under the mean reversion of
Section~\ref{sec:price-formation} is lower than a daily volatility scaled up.
The put is worth \$0.388156 per accelerator-hour. The note sells
$N/K = 390{,}625$ accelerator-hours of put \emph{notional}, the quantity the
option is written over, so the option leg is worth \$151{,}623. The fair coupon
is \Notecoupon{}\% a year, of which 4.08 points are the time value of money and
\Noteput{} points the put premium. As volatility approaches zero the put
becomes worthless and the coupon falls to 4.081\%, the one-year simple-interest
equivalent of a 4\% continuously compounded rate. Put--call parity on the
forward holds to rounding, and a Monte Carlo evaluation of
Equation~\eqref{eq:rcpayoff} reproduces the decomposition within its sampling
error.

A coupon near 20\% is the fair price of selling a one-year put twenty per cent
out of the money on an underlying with 60\% volatility. Below the strike the
holder loses \$390{,}625 for every dollar the rate falls further, or 3.9\% of
principal for each ten cents.

The volatility is assumed, and it is the largest judgement in the example.
Section~\ref{sec:price-formation} measured the two rental indices. At the
posted-price index's \GRPIsigmaCalendar{}\% a year the note pays
\NotecouponCalendar{}\%, of which \NoteputCalendar{} points are put premium: a
strike twenty per cent below spot is several standard deviations away, and the
enhancement all but vanishes. At the transaction index's
\OCPIsigmaHorizon{}\% a year over \HorizonDays-day changes the coupon is
\NotecouponOCPIHorizon{}\%, with \NoteputOCPIHorizon{} points of premium, and
at its daily \OCPIsigmaCalendar{}\% it is \NotecouponOCPIDay{}\%, with
\NoteputOCPIDay{} points. The measured range spans the product's economics from
end to end. Two further considerations bear on where the right figure lies,
though neither says how far: an option is priced on implied volatility, which
in developed commodity markets stands above realised volatility, and the note
pays on the rate a year out, a horizon longer than either series.

\begin{figure}[htbp]
\centering
\includegraphics[width=\textwidth]{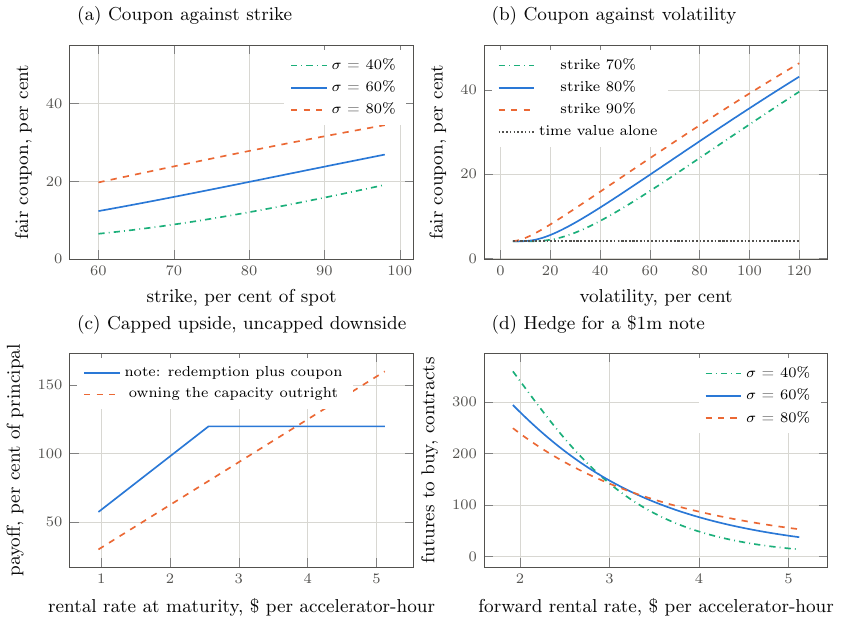}
\caption{The worked example, moved off its base case. All four panels use the
same forward rate, term and risk-free rate as the text, so the base case,
strike at 80\% of spot and volatility 60\%, appears in each. Panel~(a): the fair
coupon of Equation~\eqref{eq:couponsolve} against the strike, for three
volatilities. Panel~(b): the same coupon against volatility, for three strikes;
as volatility vanishes the coupon falls to its time-value term, the horizontal
line. Panel~(c): the redemption of Equation~\eqref{eq:rcpayoff} plus the
coupon, against owning the exposure outright; upside stops at par plus the
coupon and downside does not stop. Panel~(d): futures a dealer that has issued
the notes must buy to offset the rate exposure of a \$1~million issue, against
the forward rate.}
\label{fig:note}
\end{figure}

Figure~\ref{fig:note} moves each input separately. The coupon is close to
linear in the strike over the range a counterparty would ask for, so a
negotiation over the protection level is close to a negotiation over price. It
is not linear in volatility: the curve is flat near zero and then steepens, so
a modest error in the volatility assumption is a large error in the coupon.

\subsection{Who takes each side}

The issuer is long the put, which is what the operators of
Section~\ref{sec:deals} need. An operator with fixed debt service against
floating rental income becomes insolvent at some rental level; a put struck
near that level converts that exposure into a known premium. The natural
issuers are therefore the leveraged operators and the vehicles financing them,
a narrower and weaker population than ``sellers of compute'' suggests.

The natural holders are investors paid to take the risk: buyers of high-yield
credit, for whom the note is an alternative to the operators' own bonds, and
long-horizon funds seeking income and exposure to AI infrastructure. Firms
across the ecosystem may also hold it to support the operators they trade with, as NVIDIA and
Meta already do by agreeing to buy unsold capacity from CoreWeave and Nebius
(Section~\ref{sec:mechanism}); that is commercial support, not a hedge. Lenders
to operators are not natural holders: they already hold this exposure
(Section~\ref{sec:deals}) and are natural buyers of the protection.

The holder is short the put and earns the coupon. This class of product has a
long record of being mis-described to buyers~\citep{finra1009}, so the risk
must be stated exactly: at $R_T = 0$ the holder receives nothing, so the
maximum loss is the whole principal, while the gain is capped at par plus the
coupon. What the note offers against owning the exposure outright is that the
first twenty per cent of decline costs nothing and that the compensation is
contractual.

\subsection{Collateral}

A reverse convertible is ordinarily an unsecured claim on its issuer, and that
will not do here, though not because of wrong-way risk. Under
Equation~\eqref{eq:rcpayoff} the issuer's obligation \emph{falls} as the
rental rate falls; it owes par only when its own business is healthy. What
remains is ordinary credit risk on a leveraged counterparty, and it is large:
the holder has $N(1+c)$ at risk, about \$1.199 million on \$1 million in the
worked example. Collateral covering only the shortfall below the strike would
not address it, because that is the amount the issuer does \emph{not} owe. The
note therefore needs collateral covering par plus coupon, held in an asset
uncorrelated with the rental rate, in a vehicle bankruptcy-remote from the
operator, for the life of the note. That cuts against the client base: an
operator whose capital is committed to hardware and debt service may have none
to post. Whether a third party can supply the collateral, and at what cost, is
open.

\subsection{What can be hedged on an exchange}

An operator may issue the note itself, or a dealer may stand between the two
sides: the dealer issues notes to holders, which leaves it long the puts, and
sells the same protection to operators, which leaves it short. What it has not
matched it must hedge. A dealer that has issued the worked example's notes and
not yet sold the protection on is long 390{,}625 accelerator-hours of put
notional.
The put's \emph{delta}, the change in its value for a one-dollar change in the
forward rate, is $-0.2410$, so the dealer's position is short 94{,}136
accelerator-hours of forward: it loses as the rate rises, and it flattens by
\emph{buying} futures. At 730 accelerator-hours per CME contract that is 129
contracts, against 535 for the full notional. Panel~(d) of
Figure~\ref{fig:note} shows the hedge growing as the rate falls towards the
strike.

Tenor is the first mismatch. The largest contracted books run take-or-pay for
up to fifteen years (Section~\ref{sec:mechanism}); if their prices are fixed,
the residual exposure sits at renewal. CME's specification runs monthly out to thirty-six
months. CoreWeave's DDTL~5.5 finances about two years beyond contracts
averaging three (Section~\ref{sec:ddtl}), so its renewal tail begins about where
the listed curve ends; Applied Digital's fifteen-year terms place its renewals
a decade beyond it. The note can be written to either horizon; the futures
curve cannot.

The rest cannot be hedged on an exchange. No option on any compute index is
listed, so the \emph{gamma} and \emph{vega} of the puts, their sensitivities to
moves in the rate and in volatility, must be carried or crossed against other
underlyings. A jump through the strike between rebalancing dates cannot be
replicated by any delta hedge in a monthly-settled future, and for whoever is
short the puts, the holder or a dealer that has sold protection to an
operator, it is the dominant risk. On CME and ICE, accelerator models other
than H100 and B200 are hedged only by proxy.

\subsection{Credit as the other hedge}

Whoever is short the puts, the holder or a dealer that has sold protection to
an operator, loses when the rental rate falls. The operators whose solvency the
rental rate governs are the issuers that credit default swaps reference, so a
fall large enough to take the note through its strike should widen their
spreads, and protection on them is a candidate hedge. Let $H$ denote a candidate hedge
instrument, $\Delta R$ and $\Delta H$ the changes in the rental rate and in $H$
over the same interval, and $\rho$ their correlation. The variance-minimising
hedge ratio is
\begin{equation}
\label{eq:hedgeratio}
h^{*} \;=\; \frac{\operatorname{Cov}(\Delta R, \Delta H)}{\operatorname{Var}(\Delta H)},
\end{equation}
and the variance that survives it is
\begin{equation}
\label{eq:residvar}
\Sigma \;=\; \operatorname{Var}(\Delta R)\,\bigl(1 - \rho^{2}\bigr)
\end{equation}
\citep[sec.~3.4]{hull2015}. Equation~\eqref{eq:hedgeratio} has not, to our
knowledge, been evaluated with a credit instrument as $H$, because we found no
measurement of the relationship (Section~\ref{sec:deals}). Until somebody does, a credit hedge is
a position on a believed relationship and should be sized as one. The ordering
by revenue dependence on the artificial intelligence ecosystem in
Section~\ref{sec:landscape} is a prior over which names to test, not an
estimate of $h^{*}$.

Three features of the available instruments are already clear. The liquidity
sits in the wrong name: Oracle has the only deep single-name market, and
protection on Oracle hedges the financing cycle rather than the rental rate.
Nebius, which sells accelerator capacity directly, and Applied Digital, the
most dependent name in the screen of Section~\ref{sec:landscape}, have no
single-name contract that we could establish. And a concentrated exposure is a small share
of a broad credit index, so Equation~\eqref{eq:hedgeratio} returns a small
ratio against the index and Equation~\eqref{eq:residvar} leaves nearly all the
variance.

One instrument is relevant without any correlation assumption. Being short
puts is being short volatility, and the convertible notes these issuers have sold carry
traded volatility on their equity, hedged by dealers who say so in the offering
documents. It is a different underlying with every basis problem described
above, but it is the only place in this market where an option position can be
put on at size.

\subsection{What we cannot yet price}

No series yet measures the volatility the note needs. The posted-price index
spans accelerator families, and the transaction index for a single accelerator
runs \OCPIdays{} days, enough to show that its daily moves largely reverse but
not to measure variance over a year. A volatility would not be enough in any
case: with the scheduled jumps of Equation~\eqref{eq:process} the market is
incomplete~\citep[sec.~6.9.1]{tsay2010}, and the premium for jump risk cannot be
derived from the underlying. It is ordinarily recovered from listed option
prices, and there are none. Until options list, any price struck on this note
embeds a judgement about jump risk that can be stated but not calibrated. The
residual variance of Equation~\eqref{eq:residvar}, by contrast, is measurable
as soon as a rental series of sufficient length exists, and it bounds how much
of the note a dealer can carry before any option model is chosen.

\section{Conclusion}
\label{sec:conclusion}

Compute and inference are priced separately, and the evidence assembled here
says they must be hedged separately. Every exchange contract on compute filed,
announced or listed is written on accelerator capacity, while an inference
buyer's exposure runs through a token price that moves with throughput, margin
and the mix of models as well as with the rental rate. The rental rates the CME and ICE
contracts settle on are the visible residual of a market whose economics sit in
undisclosed multi-year contracts.

On the supply side, the exposure that needs hedging already sits in credit
agreements, in the years by which operators' debt outlives their contracted
revenue. Disclosures do not allow that quantity to be measured, and the one
financing that prices it settles the risk by prepayment against a private model
rather than by a hedge. A reverse convertible note would let the operators
holding that exposure buy protection, at a coupon set by a volatility no public
series yet measures.

Four measurements would change what can be said, and each becomes possible as
the market matures. The variance of the rental rate over a year, on a
single-accelerator index and under the settlement convention of a listed
contract, sets the price of every option in Section~\ref{sec:products}. The
price mechanism inside operators' committed contracts decides whether their
residual exposure sits within the contract or at renewal. The relationship
between rental rates and the credit spreads of the operators borrowing against
them is computable from published series and has not been measured. And the
number of tokens a task consumes, which no public source reports, decides
whether a falling price per token is a falling cost of work. Once a listed
contract reaches reportable size, the regulator's weekly positioning report
will add the first direct observation of who hedges, and on which side.

\bibliographystyle{plainnat}
\bibliography{refs}

\end{document}